\documentclass[amsfonts,aps,physrev,twocolumn,superscriptaddress]{revtex4-2}

\usepackage{dcolumn}
\usepackage{float} 
\usepackage{bm} 
\usepackage{braket} 
\usepackage{graphicx} 
\usepackage{hyperref} 
\usepackage{xcolor} 
\usepackage{amsmath} 
\usepackage{cleveref} 

\begin{document}

\title{Electronically-controlled one- and two-qubit gates for transmon quasicharge qubits}

\author{Nicholas M. Christopher}

\author{Deniz E. Stiegemann}

\affiliation{ARC Centre for Engineered Quantum Systems, School of Mathematics and Physics, University of Queensland, Brisbane, Queensland 4072, Australia}

\author{Abhijeet Alase}

\affiliation{ARC Centre for Engineered Quantum Systems, School of Physics, University of Sydney, New South Wales 2006, Australia}
\affiliation{Department of Physics, Concordia University, Montreal, QC H4B 1R6, Canada}

\author{Thomas M. Stace}
\email[]{stace@physics.uq.edu.au}

\affiliation{ARC Centre for Engineered Quantum Systems, School of Mathematics and Physics, University of Queensland, Brisbane, Queensland 4072, Australia}

\date{\today}

\begin{abstract}
Superconducting protected qubits aim to achieve 
sufficiently low error rates so as 
to allow realization of error-corrected, utility-scale 
quantum computers. 
A recent proposal encodes a protected qubit
in the quasicharge degree of freedom
of the conventional transmon device.
Operating such a protected
 `quasicharge qubit' requires implementing new strategies. 
Here we show that an electronically-controllable
tunnel junction formed by two topological superconductors
can be used to implement single- and two-qubit 
gates on quasicharge qubits. Schemes for both these gates
are based on the same dynamical $4\pi$-periodic Josephson effect 
and therefore have the same  gate times and error characteristics. 
We simulate the dynamics of 
a topological Josephson junction
in a parameter regime with non-negligible charging energy, and characterize the robustness of such gate operations against charge noise. 
Our results point to a compelling strategy for 
implementation of quasicharge qubit gates 
based on junctions of minimal Kitaev chains 
of quantum dots.

\end{abstract}

\maketitle

\section{\label{sec:intro} Introduction}

Building a utility-scale quantum computer requires
qubit architectures that support operations 
with extremely low error rate. In particular,
the error rate per operation needs to be well 
below `threshold' rates for the error-correction
schemes to be effective. The best
contemporary qubit architectures,
including the transmon~\cite{koch_charge-insensitive_2007},
currently achieve error rates 
comparable to the threshold rates,
and there is a need to improve this further.

Several superconducting qubit designs have been proposed 
with varying levels of resilience to noise sources, 
including the $0$-$\pi$ \cite{brooks_protected_2013} and fluxonium \cite{manucharyan_fluxonium_2009} qubits.
Control of these `protected' qubits is an area of active study
\cite{kolesnikow_protected_2025}.
The current work outlines a control scheme for
a recently proposed protected qubit.

Motivated by the promise of protected qubits,
a recent work~\cite{thanh_le_building_2020} 
proposes encoding of a qubit in the
quasicharge degrees of freedom in the conventional
transmon device, which we refer to as a `quasicharge qubit'. 
While the qubit itself can reside in the 
conventional transmon circuit,
implementing gate operations on this qubit 
was anticipated to require additional
components in the circuit. Specifically, Ref.~\cite{thanh_le_building_2020}
shows that a $4\pi$-junction circuit element, 
which contributes a $\cos(\hat{\phi}/2)$ term to the Hamiltonian, 
would allow accurate qubit gate operations.
The aim of this work is
to show that a junction formed by topological superconductors,
as suggested in Ref.~\cite{thanh_le_building_2020},
suffices for this purpose, and to assess the performance of
the resulting schemes for gate implementation.

The phenomenology of a $4\pi$-junction
is based on a theoretical proposal that
topological superconductors exhibit a 
robust $4\pi$-periodic `fractional' 
Josephson effect~\cite{kitaev_unpaired_2001}.
The origin of this effect is attributed to the
presence of Majorana zero-energy modes on either
side of the junction, which allow coherent transfer of
half of a Cooper-pair.
This contrasts the Josephson effect in
topologically trivial superconductors 
in which charge is transferred across a Josephson junction
in integer numbers of Cooper pairs.
Systems involving topological superconducting junctions
have been studied in the literature in various contexts 
\cite{keselman_spectral_2019, ginossar_microwave_2014, avila_majorana_2020, lutchyn_majorana_2010, pino_minimal_2024}
(see \cite{prada_andreev_2020} for a recent review).
However, most existing work emphasizes the spectroscopy of the system
for the purposes of detecting signatures of topological superconductivity.
Here we consider the time-domain dynamics
of one or more transmons coupled to a microscopic model of a $4\pi$ element
to evaluate their potential utility as physically-motivated,
electronically-controlled quantum gates.

When a topological superconducting junction is added in
parallel to a transmon circuit, the superconducting degrees of freedom
become correlated with the junction degrees of freedom,
allowing the fermion parity of the superconducting island to change.
The even and odd parity states
of the superconducting island
form the computational basis states of the quasicharge qubit. 
Such a qubit has been 
studied independently in 
other works~\cite{ginossar_microwave_2014},
where it was called a Majorana-transmon
(MT) qubit~\cite{ginossar_microwave_2014}.
RF-controlled gates
have been proposed~\cite{yavilberg_fermion_2015} for these qubits.

The main result of this paper is 
the numerical demonstration of single- and two-qubit gates
on the quasicharge qubit
via electronic control of the junction tunneling strength.
Furthermore, we quantify the effect of charge noise
on these gate schemes
as a function of system and noise parameters.
We also show that,
under some reasonable assumptions,
the error rate due to charge noise
remains constant with respect to the length
of the topological superconductor.

Although simulating the full fermionic Hilbert space
is exponentially expensive with respect to the length of the topological superconductor, 
a minimal two-site model of the topological superconductor
suffices for our purpose
-- enabling a simulation of
single- and two-qubit quantum gates.
We use a set of two minimal Kitaev chains 
(each with two fermion sites)
as a model of the $4\pi$-junction element.
We then simulate a transmon
coupled to the junction
and find coherent transitions
between the two computational states.
We further show that connecting two transmons 
in series with the junction
applies an entangling two-qubit gate. 

This work provides an alternative perspective on Majorana-transmon qubits~\cite{ginossar_microwave_2014}
as a standard transmon in an extended Hilbert space~\cite{thanh_le_building_2020},
enabled via coupling to the $4\pi$-periodic inductive element
realized by the topological junction.
Further,
it builds towards a superconducting quantum computing architecture
controlled using only DC signals.
Another feature of our scheme for gate operation
is that one- and two-qubit gates rely on
an identical physical mechanism
and thus have similar operation speeds.
In comparison,
the standard for two-qubit entangling gates for traditional transmons
is two orders of magnitude slower than the fastest single-qubits gates
implemented through microwave driving \cite{kandala_demonstration_2021}.

The organization of the rest of this paper is as follows.
In \cref{sec:background} we review the standard transmon
and see how extending the Hilbert space allows for an alternative qubit encoding.
We then review Kitaev's model of a $4\pi$-periodic element
and some potential experimental realizations.
Next, we couple the transmon and the $4\pi$-periodic element
to define the system of interest,
known as the Majorana-transmon (MT) qubit \cite{ginossar_microwave_2014}.
In \cref{sec:unitary-dynamics} we simulate the unitary dynamics
of a minimal model for the MT qubit 
tuned to a parameter `sweet spot',
and see coherent Rabi cycles
when the chains are initialised in the appropriate state.
Motived by these simulations,
we derive a simple model by projecting the full system into a qubit subspace.
In \cref{sec:noisy-dynamics}, we simulate the effect of charge noise
on this minimal junction.
Then, we derive an expression for the leakage out of the qubit subspace
as a function of the length of the Kitaev chains in perturbation theory
and find that the results of our minimal model are robust for longer chains
and parameter regimes away from the fine-tuned sweet spot.
Finally, in \cref{sec:two-qubit},
we extend our analysis to a two-qubit design
which we demonstrate can be used to perform entangling gates using the same electronically-controlled, fractional Josephson effect used for single-qubit gates.

\section{\label{sec:background}Background}

Following Ref \cite{thanh_le_building_2020}, we explain how an alternative qubit may be encoded in a transmon system in which the Hilbert space is extended to include states with a periodicity greater than $2\pi$ in the superconducting phase difference \newcommand{\scphase}{\phi}$\scphase$. This qubit is expected to be intrinsically resistant to dephasing and relaxation, making it a protected qubit. Next, we review the Kitaev chain model of a one-dimensional topological superconductor and present the fractional Josephson effect in topological superconducting junctions which realizes a $4\pi$-periodic superconducting circuit element. We briefly discuss recent realizations of minimal Kitaev chains. Finally, we put the pieces together to define the Majorana-transmon qubit as a device in which a transmon is coupled to a $4\pi$-periodic element.

\subsection{\label{sec:transmon}The transmon qubit}

\begin{figure}
    \centering
    \includegraphics{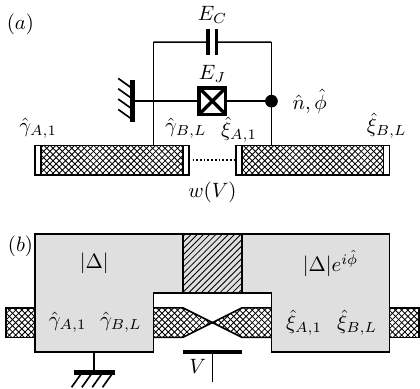}
    \caption{\label{fig:circuit} (a) A transmon (with degrees of freedom $\hat{n}$ and $\hat{\phi}$) consisting of a capacitor of charging energy $E_{C}$ in parallel with a Josephson junction with Josephson energy $E_{J}$ is coupled in parallel to a topological superconducting junction (crosshatched) that supports Majorana zero modes $\hat{\gamma}_{I,j}$ and $\hat{\xi}_{I,j}$ (open rectangles) on the boundaries between the topological and the trivial phases. (b) A potential physical setup realizing the circuit of (a). Each of the two Kitaev chains (crosshatched) is proximity coupled to the superconducting islands on the left and right -- each hosting Majorana zero modes $\hat{\gamma}_{I,j}$ and $\hat{\xi}_{I,j}$ at its ends. The gray region represents a bulk superconductor. The pairing amplitude $|\Delta|$ is related to the bulk superconducting gap. A Josephson junction (single-hatched) with gauge-invariant phase difference $\hat{\phi}$ couples the floating superconducting island to the grounded superconductor, defining a transmon. The ``twist" in the Kitaev chains indicates the tunable weak link. A gate voltage $V$ controls the tunneling potential $w$ between the two sides of the wire.}
\end{figure}

The transmon qubit \cite{koch_charge-insensitive_2007} has become ubiquitous in the design of large-scale quantum computers \cite{acharya_quantum_2025, kim_evidence_2023}. The transmon circuit consists of a Josephson junction storing a Josephson energy \newcommand{\jenergy}{E_{J}}$\jenergy$ in parallel with a capacitor of charging energy \newcommand{\cenergy}{E_{C}}$\cenergy$ (\cref{fig:circuit}). The coordinates of this circuit are taken to be the branch charge of the Josephson junction measured in Cooper pair number \newcommand{\sccharge}{n}$\hat{\sccharge}$ and the branch flux across the Josephson junction $\hat{\scphase}$, in units of flux quantum, which is identified with the gauge-invariant superconducting phase difference across the junction. These coordinates are taken to be canonically conjugate, $[\hat{\scphase}, \hat{\sccharge}] = i$.
The Hamiltonian representing the circuit is \newcommand{\htransmon}{\hat{H}_{\rm T}}
\begin{equation}
\htransmon = \cenergy \hat{\sccharge}^{2} - \jenergy \cos \hat{\scphase}.
\label{eq:hamiltonian-transmon}
\end{equation}

The non-linear inductance of the Josephson junction produces the anharmonic spectrum required to define a qubit using the lowest two energy levels of $\htransmon$.

The defining parameter regime of the transmon circuit is $\jenergy / \cenergy \gtrsim 50$ \cite{koch_charge-insensitive_2007}. In this regime, the transmon is intrinsically resilient to certain types of errors induced by charge noise while the spectrum remains sufficiently anharmonic.

An intuitive way to understand this resilience to certain errors is to note that the Josephson potential term in \cref{eq:hamiltonian-transmon} couples charge eigenstates $\ket{n}$. This means that, in the transmon regime, the energy eigenstates of system will have support over many charge states. This implies that a charge measurement (in the form of environmental charge noise, for example) cannot sharply determine the energy of the system and so the phase coherence of the state is protected.

While the transmon is protected against pure dephasing due to charge noise, it has limited protection from relaxation. To build a qubit that is simultaneously protected against dephasing and relaxation, we must find an alternative to the transmon qubit~\cite{gyenis_moving_2021}. Ref. \cite{thanh_le_building_2020} shows that a protected qubit may be constructed by retaining the transmon Hamiltonian \cref{eq:hamiltonian-transmon} while extending its Hilbert space to accommodate states of periodicity in $\scphase$ greater than $2\pi$.

\subsection{\label{sec:bigger-hilbert-space}A bigger Hilbert space for a protected qubit}

By extending the Hilbert space of a transmon qubit, it was shown in Ref. \cite{thanh_le_building_2020} that a new qubit encoding is possible. This qubit is protected from both dephasing and relaxation in between gate operations. 

The traditional perspective on the superconducting phase operator \newcommand{\transmonphase}{\hat{\scphase}}$\transmonphase$ is that its spectrum is a compact interval of length $2\pi$ and thus, the phase eigenstates $\ket{\scphase}$ and $\ket{\scphase + 2\pi}$ are equivalent \cite{koch_charge-insensitive_2007}. In contrast, the perspective in which this new qubit encoding was formulated was one in which the superconducting phase operator $\transmonphase$ is taken to have the entire real line as its spectrum and thus the states $\ket{\scphase}$ and $\ket{\scphase + 2\pi}$ are orthogonal \cite{thanh_le_building_2020, le_doubly_2019, koch_charging_2009, liao_circuit_2025}. This perspective is part of a broader discussion on the support of the operator $\transmonphase$ \cite{susskind_quantum_1964, pegg_phase_1989, likharev_theory_1985}. We emphasize that, while this perspective provides a useful conceptual framework for discussing the gate operations in this work, we need not assume that the spectrum of $\transmonphase$ is non-compact. Here, we take a less dramatic approach to extending the Hilbert space. The coupling of the transmon to the Kitaev chain junction implies the existence of additional $4\pi$-periodic states in the Hilbert space (this is discussed in detail in Section \cref{sec:topological-transmon}). Nonetheless, we will embed these allowed states in the larger Hilbert space in which $\transmonphase$ has non-compact spectrum, as this perspective provides insights into the protected nature of the qubit.

When the restriction of $2\pi$-periodicity on eigenstates of $\transmonphase$ is lifted, the Josephson potential term in $\htransmon$ becomes an infinitely extended periodic potential. The spectrum of $\htransmon$ thus exhibits a band structure and a Bloch quasicharge degree of freedom \newcommand{\blochqc}{\kappa}$\blochqc$ emerges. The lowest two bands, $E_{0}(\blochqc)$ and $E_{1}(\kappa)$, are plotted for the case when $\jenergy/\cenergy = 1$ in \cref{fig:band-structure}. The states at the center of the Brillouin zone in each band correspond to the traditional transmon eigenstates while all eigenstates with $\blochqc \neq 0$ are outside of the traditional $2\pi$-periodic Hilbert space \cite{thanh_le_building_2020}.

\begin{figure}
    \centering
    \includegraphics{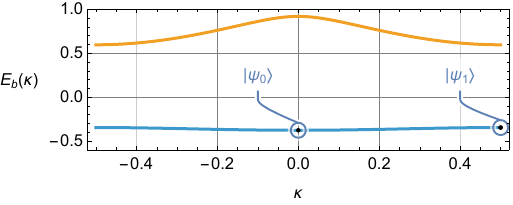}
    \caption{The first two energy bands of $\htransmon$ in an extended Hilbert space in which the restriction on phase eigenstates to be $2\pi$-periodic is lifted (based on a similar figure in \cite{thanh_le_building_2020}). The energy levels of the traditional transmon are at the points $\blochqc=0$ in each band. The states proposed by Ref \cite{thanh_le_building_2020} to encode a protected qubit, $\ket{\psi_{0}}$ and $\ket{\psi_{1}}$, are at the centre and the edge of the Brillouin zone in the first band. This plot was generated for the parameters $\jenergy/\cenergy = 1$ which is distinct from the transmon regime. The lower band is essentially flat for large $\jenergy/\cenergy$ ratios.}
    \label{fig:band-structure}
\end{figure}

The eigenstates with $\blochqc = 0$ are $2\pi$-periodic in $\scphase$, as expected, while the eigenstates with $\blochqc = 1/2$ are $2\pi$-antiperiodic. These boundary conditions appear often in the study of topological superconducting junctions \cite{keselman_spectral_2019, ginossar_microwave_2014, fu_electron_2010}. We discuss the connection between this picture and topological superconductivity in \cref{sec:discussion}.

Ref. \cite{thanh_le_building_2020} defines a computational basis for the quasicharge qubit consisting of two states in the first Bloch band. Logical $0$ is the true ground state with $\kappa = 0$ and logical $1$ is at the edge of the Brillouin zone with $\kappa = 1/2$. We will refer to these states as $\ket{\psi_{0}}$ and $\ket{\psi_{1}}$, respectively (see \cref{fig:band-structure}). For this to be a useful qubit, a process that couples these two states is required. These two states are dynamically superposed by adding a $4\pi$-periodic potential term to the Hamiltonian of the form $E_{L}\cos(\transmonphase/2)$ \cite{thanh_le_building_2020}. However, a physical system that effects such a potential term is still required for practical implementation.

The protection of the qubit from dephasing due to charge noise is inherited from the traditional transmon. Additionally, this qubit is also protected from relaxation while the gate is not operating. This is most simply seen by noting that the matrix element $\bra{\psi_{0}} \hat{\mathcal{O}} \ket{\psi_{1}}$ can only be non-zero for operators $\hat{\mathcal{O}}$ that break the $2\pi$-periodicity of the Josephson potential term. In particular, the matrix element $\bra{\psi_{0}} \hat{\sccharge} \ket{\psi_{1}}$ vanishes, implying that relaxation does not occur due to charge noise. Whether or not this protection from relaxation persists during gate operation is a question for future work.

\subsection{The Kitaev chain model of topological superconductivity}

The traditional Josephson effect leads to a $2\pi$-periodic $\jenergy \cos \hat{\scphase}$ term in the circuit Hamiltonian, \cref{eq:hamiltonian-transmon}. In contrast, a fractional Josephson effect leads to a $4\pi$-periodic term proportional to $\cos(\hat{\scphase}/2)$. The possibility of a fractional Josephson effect in topological superconductivity was discussed by Kitaev in a simple model of a one-dimensional topological superconductor \cite{kitaev_unpaired_2001} (and since then, the fractional Josephson effect has been shown to persist in more general systems exhibiting topological superconductivity \cite{fu_josephson_2009}). In this section, we review the Kitaev chain model for a one-dimensional topological superconductor. In the next section, we explain how the Kitaev chain is used to model a $4\pi$-periodic topological superconducting junction.

The Kitaev chain is a simple model of a topological superconductor. The model consists of \newcommand{\chainlength}{L}$\chainlength$ discrete, spinless fermion sites described by creation and annihilation operators \newcommand{\leftchainfermion}[1]{\hat{b}_{#1}}$\leftchainfermion{j}$ and $\leftchainfermion{j}^{\dagger}$ which satisfy the canonical anticommutations relations $\{\leftchainfermion{j},\leftchainfermion{k}^{\dagger}\} = \delta_{jk}$.

The model is specified by three parameters: an on-site potential given by a real-valued \newcommand{\onsite}{\mu}$\onsite$, a nearest-neighbour hopping potential given by positive \newcommand{\hopping}{t}$\hopping$ and a complex, nearest-neighbour pairing potential \newcommand{\pairing}{\Delta}$\pairing = |\pairing| e^{-i \theta}$. In terms of these parameters, the Hamiltonian describing the Kitaev chain is \newcommand{\kitaevhamiltonian}[1]{\hat{H}_{\rm K}^{#1}}
\begin{equation}
\kitaevhamiltonian{} = \sum_{j = 1}^{\chainlength} \big( -\frac{\mu}{2}\leftchainfermion{j}^{\dagger} \leftchainfermion{j} - t \leftchainfermion{j}^{\dagger} \leftchainfermion{j + 1} + \Delta \leftchainfermion{j}^{\dagger} \leftchainfermion{j + 1}^{\dagger} + h.c. \big).
\label{eq:hamiltonian-kiteav-chain}
\end{equation}
This model exhibits two distinct topological phases, 
theoretically distinguished through the bulk topological
invariants such as the sign of the Pfaffian~\cite{kitaev_unpaired_2001} 
or other equivalent forms~\cite{alase2023wiener}. The trivial phase occurs when $|\onsite| > 2t$ and the non-trivial phase, in which the $4\pi$ Josephson effect emerges, occurs when $|\onsite| < 2t$.

The topological phase is characterized by the appearance of sub-gap bound states, exponentially localized at the ends of the chain, known as Majorana zero modes. The size of these modes is characterized by the Majorana localization length $\ell_{0}$ \cite{alase_erasure_2024}.

A parameter range within the topological phase which greatly simplifies analysis is $\onsite = 0$, $\hopping = \pairing$ which we refer to as the `sweet spot'. At the sweet spot, $\ell_{0} = 0$ so that Majorana zero modes are perfectly local to the ends of the chain. Away from this sweet spot, the finite length of the chain implies there is some residual overlap between the Majorana modes on either end which is exponentially suppressed in $L / \ell_{0}$. If this overlap is too large, the topological properties of the Majorana modes, including the fractional Josephson effect, are destroyed.

To ensure the Majorana modes remain separated, we must work in a regime where either $L \gg \ell_{0}$ or, at the sweet spot where $\ell_{0}$ vanishes. The numerical simulations in this work are performed for $L = 2$ and thus, we must work at the sweet spot to ensure separation of the Majorana modes.

Some devices that attempt to produce topological superconductivity are proximized superconducting nanowires \cite{oreg_helical_2010}, Yu-Shiba-Rusinov states in chains of magnetic atoms \cite{nadj-perge_proposal_2013} and quantum dots with Rashba spin-orbit coupling \cite{dvir_realization_2023} (see \cite{flensberg_engineered_2021} for a review). 

The experiment most closely resembling the Kitaev chain model used here are the quantum dots of Ref.~\cite{dvir_realization_2023} which realize a minimal Kitaev chain system consisting of two fermion sites. They demonstrated that this setup may be tuned close to the sweet spot $\onsite = 0$, $\hopping = |\pairing|$ where topological superconductivity is expected to occur in a two-site system. One may view the generic physical setup of our device shown in \cref{fig:circuit}b as an example of such a setup. However, we do not assume a specific implementation in our analysis.

\subsection{\label{sec:fractional-josephson}The fractional Josephson Effect}

When in the topological phase, a junction composed of two Kiteav chains is expected to exhibit the fractional Josephson effect, in which a component of the Josephson current across the junction shows a $4\pi$-periodicity in $\phi$. Here, we follow the exposition of \cite{alicea_new_2012}.

Take a superconducting junction consisting of two Kitaev chains of length $\chainlength$, both with the same onsite potential $\onsite$ and hopping amplitude $\hopping$ but, with different pairing potentials $\pairing^{(l)}$ and $\pairing^{(r)}$. We will take these pairing potentials to have the same modulus with differing phases, $\pairing^{(l)} = |\pairing|e^{ i \theta_{r}}$ and $\pairing^{(r)} = |\pairing| e^{i \theta_{l}}$. Each of the chains is described by a Hamiltonian of the same form as \cref{eq:hamiltonian-kiteav-chain}. The Hamiltonian for the left chain is $\kitaevhamiltonian{(l)}$ and the Hamiltonian for the right chain is $\kitaevhamiltonian{(r)}$. We retain the fermion creation and annihilation operators $\leftchainfermion{j}$ for the left chain and use a new set of fermionic creation and annihilation operators \newcommand{\rightchainfermion}[1]{\hat{a}_{#1}}$\rightchainfermion{j}$ for the right chain.
A weak link connecting the two chains causes
coherent tunneling of single electrons
from one chain to the other,
modeled by the tunneling Hamiltonian
\newcommand{\tunnelling}{w}\newcommand{\tunnellinghamiltonian}{\hat{H}_{\tunnelling}}
\begin{equation}
\tunnellinghamiltonian = -\tunnelling (\leftchainfermion{L}^{\dagger} \rightchainfermion{1} + \rightchainfermion{1}^{\dagger} \leftchainfermion{L}),
\end{equation}
where the real parameter $w$ is the tunneling potential
(in our device, $w$ is controlled via an external gate voltage $V$).
In total, the Hamiltonian of the full junction system is \newcommand{\hjunction}{\hat{H}_{4 \pi}}
\begin{equation}
\hjunction = \kitaevhamiltonian{(l)} + \kitaevhamiltonian{(r)} + \tunnellinghamiltonian.
\label{eq:junction-hamiltonian}
\end{equation}

To see the fractional Josephson effect, we assume we are at the sweet spot and express  $\tunnellinghamiltonian$ in terms of Majorana operators. The Majorana operators for the left chain are \newcommand{\majoranaleft}[2]{\hat{\gamma}_{#1,#2}}$\majoranaleft{I}{j}$ and for the right chain are \newcommand{\majoranaright}[2]{\hat{\xi}_{#1,#2}}$\majoranaright{I}{j}$ where, the lowercase index indicates the fermion site ($j \in \{1,\ldots,L\}$) and the uppercase index enumerates the two Majorana operators associated to each site ($I \in \{A,B\}$). These operators are Hermitian and obey the algebra 
\begin{align}
\{\majoranaleft{I}{i},\majoranaleft{J}{j}\} &= 2 \delta_{I,J} \delta_{i,j} \hat{I}\\
\{\majoranaright{I}{i},\majoranaright{J}{j}\} &= 2 \delta_{I,J} \delta_{i,j} \hat{I}\\
\{\majoranaright{I}{i},\majoranaleft{J}{j}\} &\equiv 0.
\end{align}
In terms of these Majorana operators, the fermion operators in $\tunnellinghamiltonian$ are $\leftchainfermion{L} = e^{-i \theta_{l} / 2} ( \majoranaleft{B}{L} + i \majoranaleft{A}{L} ) / 2$ and \mbox{$\rightchainfermion{1} = e^{-i \theta_{r} / 2}( \majoranaright{B}{1} + i \majoranaright{A}{1}) / 2$}.

We then project onto the ground state space by only retaining the zero modes, $\majoranaleft{B}{L}$ and $\majoranaright{A}{1}$, on the ends of each chain,\newcommand{\junctionfermion}{\hat{g}_{0}} (this derivation is done in detail for our device in \cref{app:projection})
\begin{equation}
\tunnellinghamiltonian^{(0)} = -\frac{\tunnelling}{2} \cos(\theta / 2) i \majoranaleft{B}{L} \majoranaright{A}{1} = -\tunnelling \cos(\theta / 2) \junctionfermion^{\dagger} \junctionfermion.
\label{eq:tunnelling-hamiltonian-eff}
\end{equation}
Here, the fermion occupation operator is \mbox{$\junctionfermion^{\dagger} \junctionfermion = ( \hat{I} + i \majoranaleft{B}{L} \majoranaright{A}{1} ) / 2$} (we have removed an additive constant) and the gauge-invariant phase difference is $\theta = \theta_{r} - \theta_{l}$ across the junction. The fermion described by $\hat{g}_{0}$ is an Andreev bound state local to the junction (see \cref{fig:circuit}).

\begin{figure}
    \centering
    \includegraphics{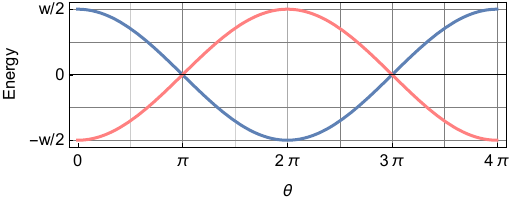}
    \caption{\label{fig:fractional-josephson} The spectrum of the effective Hamiltonian in \cref{eq:tunnelling-hamiltonian-eff} as $\theta$ is advanced by $4\pi$. The red line is the energy of the eigenstate with the junction fermion occupied while the blue is the energy of the eigenstate with the junction fermion unoccupied. The spectrum is $2 \pi$-periodic while the eigenstates are $4\pi$-periodic due to energy level crossings at $\theta = \pi, 3\pi$.}
\end{figure}

The nature of the $4\pi$-periodicity of the Josephson current arising from this term becomes clear when we consider an eigenstate of $\tunnellinghamiltonian^{0}$ when $\theta = 0$ with the junction fermion occupied and hence energy $-\tunnelling/2$ (see \cref{fig:fractional-josephson}). Under a $2 \pi$ rotation of the phase $\theta$, the state will evolve to a physically distinct state of energy $\tunnelling/2$. Similarly, an unoccupied state goes from energy $\tunnelling/2$ to energy $-\tunnelling/2$ under the same rotation. The system will only return to its initial state when the phase is rotated by a further $2 \pi$ for a total of a $4 \pi$ rotation. While the spectrum of $\tunnellinghamiltonian^{(0)}$ is still $2\pi$-periodic in $\theta$, the eigenstates become $4\pi$-periodic. This is the fractional Josephson effect (see \cref{fig:fractional-josephson}). The way in which this kinematic effect manifests in the dynamics of the system is a key result of this work.

\subsection{\label{sec:topological-transmon}Topological-transmon hybrids}

The device central to this work is a transmon circuit coupled in parallel with a $4\pi$-periodic topological junction consisting of two Kitaev chains connected by a weak link (see \cref{fig:circuit}). This same system has been studied previously in the context of detecting the existence of a topological superconducting phase and defining qubits for topological quantum computation \cite{ginossar_microwave_2014, avila_majorana_2020, keselman_spectral_2019, pino_minimal_2024, karki_physics_2024, hassler_top-transmon_2011, rodriguez-mota_revisiting_2019}. In this work, we will use the terminology of Ref. \cite{ginossar_microwave_2014} and refer to the system as a Majorana-transmon (MT) qubit. The Hamiltonian of this system is\newcommand{\hmt}{\hat{H}_{\rm MT}}
\begin{equation}
\hmt = \htransmon(\transmonphase) + \hjunction(\transmonphase).
\label{eq:mt-hamiltonian}
\end{equation}
Here, $\htransmon(\transmonphase)$ is the transmon Hamiltonian given in \cref{eq:hamiltonian-transmon} and $\hjunction(\transmonphase)$ is the Hamiltonian given in \cref{eq:junction-hamiltonian} with the gauge-invariant, superconducting phase difference $\theta$ quantized and identified with the superconducting phase difference on the Josephson junction of the transmon $\transmonphase$. We take the pairing amplitudes on the Kitaev chains to be $\pairing^{(l)} = |\pairing|$ and $\hat{\pairing}^{(r)} = e^{i \transmonphase} |\pairing|$. This identification is physically motivated by assuming the superconductivity of the Kitaev chain is induced by proximity with a bulk superconductor that is used to construct the Josephson junction of the transmon (see \cref{fig:circuit}b). 

To analyze the Hamiltonian $\hmt$, it is useful to perform a change of frame \cite{ginossar_microwave_2014, keselman_spectral_2019, van_heck_coulomb_2011}. In this new frame, the only terms which couple the transmon and the Kitaev chain appear in the (transformed) tunneling Hamiltonian. This frame makes it manifest that the chains and the transmon are decoupled by setting the tunneling amplitude to zero. Such a mechanism is crucial for us to use the junction to apply controllable quantum gates.

The unitary transformation that effects this change of frame is \newcommand{\unitarytransform}{\hat{U}}\newcommand{\rightchaincharge}{\hat{n}^{(r)}}
\begin{equation}
\unitarytransform = e^{i \transmonphase \rightchaincharge / 2}.
\label{eq:change-of-frame}
\end{equation}
Here, $\rightchaincharge = \sum_{j = 1}^{\chainlength} \rightchainfermion{j}^{\dagger} \rightchainfermion{j}$ is the total number of fermions on the right Kitaev chain (the choice of the right or left chain here is a choice of the ground node in the circuit in \cref{fig:circuit}).

For a generic operator in the original frame $\hat{\mathcal{O}}$, we denote the corresponding operator in the new frame by $\tilde{\mathcal{O}} \equiv \unitarytransform \hat{\mathcal{O}} \unitarytransform^{\dagger}$. Under this change of frame, the system Hamiltonian becomes\newcommand{\htransmontilde}{\tilde{H}_{\rm T}}\newcommand{\kitaevhamiltoniantilde}[1]{\tilde{H}_{\rm K}^{#1}}\newcommand{\htunnellingtilde}{\tilde{H}_{w}(\transmonphase)}\newcommand{\hjunctiontilde}{\tilde{H}_{4 \pi}(\transmonphase)} (see \cref{app:unitary-transform} for the details of the transformation) \newcommand{\hmttilde}{\tilde{H}_{\rm MT}}
\begin{equation}
\hmttilde = \htransmontilde(\transmonphase) + \hjunctiontilde.
\label{eq:hamiltonian-mt-new-basis}
\end{equation}
Here, $\htransmontilde$ is given by
\begin{equation}
\htransmontilde = \cenergy (\hat{\sccharge} - \rightchaincharge / 2)^2  - \jenergy \cos\transmonphase,
\end{equation}
and the tunneling junction $\htunnellingtilde$ is related to $\tunnellinghamiltonian$ by mapping all fermion operators on the right Kitaev chain as operators,
\begin{equation}
\rightchainfermion{j} \mapsto e^{-i \transmonphase / 2} \rightchainfermion{j}.
\end{equation}
The total effect of this transformation on $\hjunctiontilde$ is to remove any dependence on the transmon phase $\transmonphase$ from $\kitaevhamiltoniantilde{(l)}$ and $\kitaevhamiltoniantilde{(r)}$ and move it into $\htunnellingtilde$ as
\begin{equation}
\htunnellingtilde = -w (e^{-i \transmonphase / 2} \leftchainfermion{L}^{\dagger} \rightchainfermion{1} + e^{i \transmonphase / 2} \rightchainfermion{1}^{\dagger} \leftchainfermion{L}).
\end{equation}

The eigenstates $\ket{\tilde{\psi}_{i}}$ of $\hmttilde$ are related to the eigenstates $\ket{\psi_{i}}$ of $\hmt$ as
\begin{equation}
\ket{\tilde{\psi}_{i}} = \unitarytransform \ket{\psi_{i}}.
\end{equation}
In the phase basis, this relation is
\begin{equation}
\tilde{\psi}_{i, n^{(r)}}(\scphase) = e^{i \phi n^{(r)} / 2} \psi_{i}(\scphase).
\label{eq:new-boundary-condition}
\end{equation}
While the eigenstates of $\hmt$, $\psi_{i}(\scphase)$ are $2\pi$-periodic in $\phi$, \cref{eq:new-boundary-condition} shows that, in the new frame, the states $\tilde{\psi}_{i}(\phi)$ may be $2\pi$-periodic or $2\pi$-antiperiodic in $\phi$, depending on the parity of the integer $n^{(r)}$. By analogy with the discussion of \cref{sec:bigger-hilbert-space}, we expect the state $\tilde{\psi}_{0, n^{(r)}}(\phi)$ with $n^{(r)}$ even to correspond to the state at the center of the Brillouin zone in \cref{fig:band-structure} while the state with $n^{(r)}$ odd should correspond to the state at the edge of the Brillouin zone. This analogy is supported by numerical computation of the energies of these two states which confirms that $\tilde{\psi}_{0, n^{(r)}}(\phi)$ with $n^{(r)}$ odd has a higher energy that the state with $n^{(r)}$ even for $\jenergy/\cenergy = 1$. In the limit $\jenergy \gg \cenergy$ the energies of the two states are equal to within numerical error, also consistent with the band structure picture of Ref. \cite{thanh_le_building_2020}. To keep notation simple, we will denote the state with $n^{(r)}$ odd and the transmon in its ground state by $\ket{\tilde{\psi}_{1}}$ and the state with $n^{(r)}$ even and the transmon in its ground state by $\ket{\tilde{\psi}_{0}}$.

We notice in \cref{eq:new-boundary-condition} that, in the new frame, the degrees of freedom for the transmon and the Kitaev chains have been combined to form a degree of freedom which is a hybrid of both systems. This shows that, although one may hope to treat the Kitaev chain junction as an element separate from the transmon, to extend the Hilbert space of the transmon in this case, it is necessary to treat them as a hybrid system. This fact is well-appreciated in the topological-transmon hybrid literature \cite{ginossar_microwave_2014, hassler_top-transmon_2011, fu_electron_2010, keselman_spectral_2019} and we see that is also manifests in this alternative perspective.

\section{\label{sec:dynamics}Dynamics}

To demonstrate the principle behind our gate, we first perform a simulation of the system initialized in a particularly chosen state and show that the system exhibits oscillations between the two computational basis states. We then project the full system onto the subspace identified in the simulations which is approximately preserved by the dynamics. This gives us an effective two-dimensional model in which we demonstrate a protocol for performing $R_{X}$-gates. We extend this analysis to characterise the effect of charge noise on the junction gate voltage. Finally, we show that the same setup can be used to perform entangling $R_{XX}$-gates on a two-qubit system.

\subsection{\label{sec:unitary-dynamics}Unitary dynamics of the Kitaev chain junction}

We begin by simulating the dynamics of the full transmon plus Kitaev chain junction system described (after a change of frame) by \cref{eq:hamiltonian-mt-new-basis}. Simulating this system is computationally expensive for long chains. Therefore, we consider a minimal model in which each Kitaev chain consists of only two fermion sites, $L=2$, and we tune parameters to the topological sweet spot as $\mu = 0$, $t=|\Delta| \equiv w_{F}$ to ensure the Majorana zero modes are decoupled. We shall relax these conditions in \cref{sec:noisy-dynamics}.

To make this Hamiltonian easy to simulate, we perform a Jordan-Wigner transformation, replacing the $4$ fermion sites with spins described by Pauli matrices $\hat{\sigma}_{j}^{a}$ with $a \in \{x, y, z\}$ and $j = 1, ..., 4$ (see \cref{app:jordan-wigner}). The Hamiltonian now takes the simple form
\begin{align}
\tilde{H}_{\rm JW} = \tilde{H}_{\rm T}(\hat{\phi}) - w_{F} \left( \hat{\sigma}_{1}^{x} \hat{\sigma}_{2}^{x} + \hat{\sigma}_{3}^{x} \hat{\sigma}_{4}^{x} \right) \nonumber \\
+ w \left( e^{i \hat{\phi} /2} \hat{\sigma}_{2}^{+} \hat{\sigma}_{3}^{-} + e^{-i \hat{\phi} / 2} \hat{\sigma}_{2}^{-} \hat{\sigma}_{3}^{+}\right).
\label{eq:hamilonian-jordan-wigner}
\end{align}
The spin ladder operators are \mbox{$\hat{\sigma}_{j}^{-} = (\hat{\sigma}_{x} + i \hat{\sigma}_{y})/2$} and $\hat{\sigma}_{j}^{+} = (\hat{\sigma}_{j}^{-})^{\dagger}$.

After this transformation, the Hilbert space has a tensor product structure $\mathcal{H} = \mathcal{H}_{\rm T} \otimes \mathcal{H}_{\rm K}^{(l)} \otimes \mathcal{H}_{\rm K}^{(r)}$ where we have partitioned the system into subsystems for the transmon, the left Kitaev chain and the right Kitaev chain, respectively. These subsystems are for the purposes of labeling, they do not correspond to an obvious physical partitioning of the system. This is because the change of frame mixes up the degrees of freedom for the transmon and the right Kitaev chain, as discussed in \cref{sec:topological-transmon}. We use the eigenstates $\ket{\tilde{\psi}_{i}}$ of $\tilde{H}_{\rm T}$ as a basis for $\mathcal{H}_{\rm T}$. Though, it is possible to find these states analytically using the Zac basis \cite{thanh_le_building_2020}, it is simplest computationally to find these eigenstates by numerically diagonalizing $\tilde{H}_{\rm T}$ in the truncated charge basis. We take the latter approach here. For the Kitaev chains, we use eigenstates of $\hat{\sigma}_{1}^{z}\hat{\sigma}_{2}^{z}\hat{\sigma}_{3}^{z}\hat{\sigma}_{4}^{z}$ as a basis, which we label with four bits.

In our simulations, we initialise the transmon subsystem in the ground state $\ket{\tilde{\psi_{0}}}$ and we initialise the chains in the state
\begin{equation}
\ket{\Omega} = \left( \ket{\Phi}_{l} \otimes \ket{\Phi}_{r} + \ket{\Psi}_{l}  \otimes \ket{\Psi}_{r} \right) / \sqrt{2}.
\label{eq:initial-chain-state}
\end{equation}
Where $\ket{\Phi}_{a} = (\ket{00}_{a} + \ket{11}_{a})/\sqrt{2}$ and \mbox{$\ket{\Psi}_{a} = (\ket{01}_{a} + \ket{10}_{a})/\sqrt{2}$} on chain $a = l, r$. This initial state is a ground state of $\tilde{H}_{\rm JW}$ when \mbox{$w = 0$} and so, it is possible, in principle, to prepare this state in an experiment. It is also an eigenstate of the observable
\begin{equation}
\tilde{g}_{0}^{\dagger} \tilde{g}_{0} = \frac{1}{2} \left( 1 + (\leftchainfermion{2} - \leftchainfermion{2}^{\dagger}) (\rightchainfermion{1} + \rightchainfermion{1}^{\dagger}) \right).
\label{eq:quantum-number}
\end{equation}
with eigenvalue $0$. The Jordan-Wigner representation of $\tilde{g}_{0}^{\dagger} \tilde{g}_{0}$ is $(1 - \hat{\sigma}_{2}^{x} \hat{\sigma}_{3}^{x})/2$. This corresponds to an observable before the unitary transform given by
\begin{equation}
\hat{g}_{0}^{\dagger} \hat{g}_{0} = \frac{1}{2} \left( 1 + (\leftchainfermion{2} - \leftchainfermion{2}^{\dagger}) (e^{-i \transmonphase / 2} \rightchainfermion{1} + e^{i \transmonphase /2 } \rightchainfermion{1}^{\dagger}) \right).
\label{eq:quantum-number-physical}
\end{equation}

\begin{figure}
    \centering
    \includegraphics{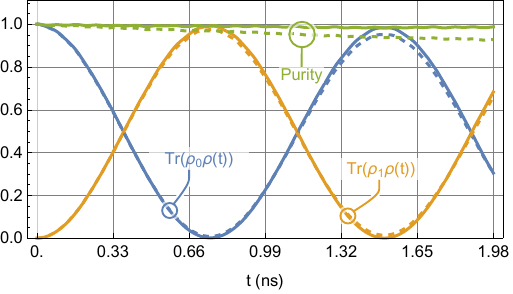}
    \caption{\label{fig:numerics} The system was initialised in the state $\rho_{0} = \ket{\tilde{\psi}_{0}}\bra{\tilde{\psi}_{0}} \otimes \ket{\Omega}\bra{\Omega}$ and evolved according to the Hamiltonian in \cref{eq:hamilonian-jordan-wigner} with $\tilde{H}_{T}(\hat{\phi})$ expressed in the charge basis with a charge truncation of $n - n^{(r)}/2 = \pm 5$. The solid blue curve is the probability for the system to be in state $\rho_{0}$ at time $t$. The solid orange curve is the probability for the system to be in state $\rho_{1} = \ket{\tilde{\psi}_{1}} \bra{\tilde{\psi}_{1}} \otimes \ket{\Omega} \bra{\Omega}$. The dashed curves indicate the same probability as the solid curve of the matching colour but for a device evolving with charge noise on the 4$\pi$-junction, modelled by \cref{eq:noisy-evolution}). This charge noise was modelled as Gaussian white-noise with power spectrum $S = \frac{0.03}{2\pi} \,\mu$eV$^{-1}$. The solid green curve is the purity of the transmon subsystem under unitary evolution while the dashed green curve is the same purity under charge noise. The physical parameters used in this simulation were $E_{J} = 1\,\mu$eV with $E_{J}/E_{C} = 200$, $w_{F} = 12\,\mu$eV \cite{dvir_realization_2023}, $w = 3\,\mu$eV.}
\end{figure}

The evolution of the state $\ket{\tilde{\psi}_{0}} \otimes \ket{\Omega}$ under $\hmttilde$ is given by the solid blue curve in \cref{fig:numerics}. Plotted as the solid yellow curve in \cref{fig:numerics} are complimentary Rabi oscillations with the state $\ket{\tilde{\psi}_{1}} \otimes \ket{\Omega}$. The purity of the transmon subsystem under this unitary evolution is plotted as the solid green curve. That the purity is essentially unity suggests that the transmon and chain subsystems, as defined above, remain in essentially a product state throughout the evolution, allowing us to treat the chain as an auxiliary system used to perform the qubit gate. In practice, this means that the quantum gate may be used repeatedly without the need to re-initialise the chains to the state $\ket{\Omega}$.

The numerical simulations demonstrate that there is a two-dimensional subspace of $\mathcal{H}$, spanned by the states $\ket{\tilde{\psi}_{0}}\otimes\ket{\Omega}$ and $\ket{\tilde{\psi}_{1}}\otimes\ket{\Omega}$, which is approximately invariant under the dynamics of $\tilde{H}_{\rm JW}$. We call this subspace $\mathcal{H}_{\rm MT}$. Projecting $\tilde{H}_{\rm JW}$ onto $\mathcal{H}_{\rm MT}$ (see \cref{app:projection}) gives the effective MT qubit model,
\begin{equation}
\tilde{H}_{\rm MT}^{(P)} =
\begin{pmatrix}
    E_{01} & (w/2) \cos(\phi/2)_{01} \\
     (w/2) \cos(\phi/2)_{10} & -E_{01}
\end{pmatrix}.
\label{eq:hamiltonian-effective-theory}
\end{equation}
Where $\cos(\phi/2)_{ij} = \bra{\tilde{\psi}_{i}} (e^{ i \hat{\phi} / 2} + e^{-i \hat{\phi} / 2}) \ket{\tilde{\psi}_{j}}/2$ and the qubit transition energy is
\begin{equation}
E_{01} = (E_{0} - E_{1})/2,
\end{equation}
in which $E_{i}$ is the energy of $\ket{\tilde{\psi}_{i}}$. As discussed in \cref{app:projection}, matrix elements of the form $\cos(\phi/2)_{ii}$ vanish in $\mathcal{H}_{\rm MT}$. Also, the off-diagonal elements appearing in \cref{eq:hamiltonian-effective-theory} may be computed by approximating $\tilde{\psi}_{i}(\phi)$ as periodic or antiperiodic sums of gaussians in the large $E_{J}/E_{C}$ limit \cite{thanh_le_building_2020} (the exact eigenstates are analysed in Ref \cite{karki_physics_2024}).

Since $\tilde{\psi}_{i}(\phi)$ is real-valued for all $i$, we may write this Hamiltonian as
\begin{align}
\tilde{H}_{\rm MT}^{(P)} = E_{01} \tilde{Z} + w(t)\cos(\phi/2)_{10} \tilde{X}/2.
\label{eq:projected-mt-hamiltonian}
\end{align}
Here, $\tilde{Z} = (\ket{\psi_{0}}\bra{\psi_{0}} - \ket{\psi_{1}}\bra{\psi_{1}}) \otimes \ket{\Omega} \bra{\Omega}$ and \mbox{$\tilde{X} = (\ket{\psi_{0}}\bra{\psi_{1}} + \ket{\psi_{1}}\bra{\psi_{0}})\otimes \ket{\Omega} \bra{\Omega}
$} are Pauli operators in the qubit subspace. We allow for time dependence in $w(t)$ to demonstrate our protocol for performing an $R_{X}$ gate. 

To perform an $R_{X}$ gate using this system, one would begin by initialising the state $\ket{\tilde{\psi}_{0}} \otimes \ket{\Omega}$ (we will not be concerned with the question of how this preparation is made). Next, we send a DC signal through the gate voltage $V$ to control the tunnelling potential $w(t)$ such that
\begin{equation}
w(t) =
\begin{cases}
    0, & t < 0 \\
    w, & 0 < t < t_{\rm gate} \\
    0, & t > t_{\rm gate}.
\end{cases}
\end{equation}
Here, the amplitude $w$ should be much larger than the qubit energy gap $E_{01}$ so that we may ignore the first, Zeeman-type term in \cref{eq:projected-mt-hamiltonian} (Alternatively, we could change to a rotating frame to remove the Zeeman-type term). This DC signal effects a unitary transformation on the MT qubit given by
\begin{equation}
\tilde{R}_{X}(t_{\rm gate}) = \exp \big[-i \frac{w}{2} \cos(\phi/2)_{10} \tilde{X} t_{\rm gate} \big],
\end{equation}
a single-qubit $R_{X}$-gate for an arbitrary rotation angle controlled by the duration of the DC pulse $t_{\rm gate}$.

\Cref{fig:numerics} shows that single-qubit $R_{X}$-gates may be applied to the MT qubit at frequencies on the order of gigahertz. This gate frequency may by tuned be controlling the amplitude of the pulsed external gate voltage $V$.

\subsection{\label{sec:noisy-dynamics}Effect of charge noise on the Kitaev chain junction}

To characterise the effect of charge noise on this gate, we first simulate the effect of this noise on our minimal model numerically. We then generalise our minimal model to longer length Kitaev chains and parameter regimes away from the sweet spot and derive an expression for the qubit leakage under charge noise in perturbation theory. 

We model the effect of charge noise on the gate voltage used to operate the qubit gate by adding a stochastic time-dependent perturbation
\begin{equation}
\delta \tilde{H}(t) = w'(t) \left( e^{-i \transmonphase / 2} \leftchainfermion{L}^{\dagger} \rightchainfermion{1} + e^{i \transmonphase /2} \rightchainfermion{1}^{\dagger} \leftchainfermion{L} \right)
\label{eq:hamiltonian-perturbing}
\end{equation}
to the Hamiltonian \cref{eq:hamiltonian-mt-new-basis}. Here, $w'(t)$ is a stationary, stochastic signal which we assume to be small compared to other relevant energies. To simulate this perturbation, we assume that $w'(t)$ is gaussian distributed white noise with zero mean and power spectrum $S(\omega) = \alpha / 2 \pi$. In this case, the master equation describing the evolution is \cite{kiely_exact_2021}
\begin{equation}
\frac{d \tilde{\rho}(t)}{d t} = -i [ \tilde{H}_{\rm JW} , \tilde{\rho}(t) ] - \frac{\alpha}{2} [ \tilde{\delta H}(t),[ \tilde{\delta H}(t), \tilde{\rho}(t) ] ].
\label{eq:noisy-evolution}
\end{equation}
We compare the noisy dynamics with $\alpha = 0.03\,\mu\text{eV}^{-1}$ to the unitary dynamics in \cref{fig:numerics}. With this noise level, the fidelity of an $R_{X}$-gate after a single Rabi cycle is $F \approx 0.975$. The fidelity may be improved by reducing the noise level $\alpha$ or running the gate at a higher speed by increasing the tunneling amplitude $w$.

While numerical simulations of the minimal model demonstrate the principle behind the gate operation, we must ensure that the error rate due to charge noise remains tolerable for more general parameter values in the topological phase. We expect the most significant source of error in these general regimes will be leakage out of the qubit subspace, particularly when more Kitaev chain sites are introduced to the model. Thus, we estimate qubit leakage by employing Fermi's golden rule to compute transition rates out of the qubit subspace $\mathcal{H}_{\rm MT}$ due to the perturbation $\delta \tilde{H}(t)$ \cite{konschelle_effects_2013,alase_decoherence_2025}.

For a general stationary, stochastic signal $w'(t)$, Fermi's golden rule gives the rate of transition between eigenstates of $\tilde{H}_{\rm MT}$, $\ket{i}$ and $\ket{f}$, as \cite{weinberg_lectures_2015}
\begin{equation}
    \Gamma_{i \to f} = 2 \pi \left|\delta H_{if}\right|^2 S(E_{f} - E_{i}),
    \label{eq:fermi-golden-rule}
\end{equation}
where $\delta H_{if}$ is the matrix element of $\delta \tilde{H}/w'$ between $\ket{i}$ and $\ket{f}$. The energies of $\ket{i}$ and $\ket{f}$ are $E_{i}$ and $E_{f}$, respectively, and $S(\omega)$ is the power spectrum of $w'(t)$. To compute the leakage rate, we take the states $\ket{i}$ in $\mathcal{H}_{\rm MT}$ (see \cref{app:leakage} for the full definition) and the states $\ket{f}$ in the orthogonal compliment $\mathcal{H}_{\rm MT}^{\perp}$ in $\mathcal{H}$. The total qubit leakage out of a given computational state $\ket{i}$ is computed by summing $\Gamma_{i\to f}$ over all states $\ket{f}$.

\begin{figure}
    \centering
    \includegraphics[width=\columnwidth]{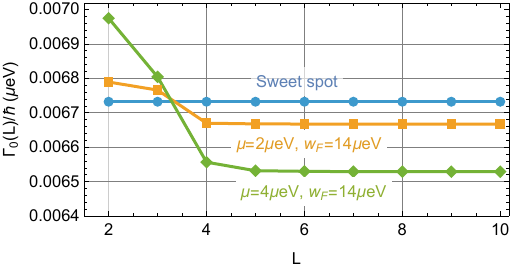}
    \caption{\label{fig:leakage} Plot of leakage rate out of $\mathcal{H}_{MT}$ with length of the Kitaev chains in the presence of Gaussian distributed white charge noise with power spectrum $S(\omega) = \frac{0.03}{2\pi} \, \mu$eV$^{-1}$. After a transient regime at small chain lengths, the leakage rate for each set of parameters becomes constant after $L = 5$. The ``sweet spot" is at $\mu = 0$, \mbox{$w_{F} = 12 \,\mu$eV}. Though shown for completeness, the orange and green curves should not be trusted for $L < 5$ (see \cref{app:leakage}).}
\end{figure}

The definitions of $\ket{i}$ and $\ket{f}$ along with the details of this computation are given in \cref{app:leakage}. The outcome is \cref{eq:leakage-full} which may be evaluated numerically to estimate the leakage rate out of states in $\mathcal{H}_{\rm MT}$ for arbitrary model parameters and noise profiles provided the noise is stationary and local to the junction. Of particular interest is how the leakage rate out of the initial state $\ket{\tilde{\psi}_{0}} \otimes \ket{\Omega}$ scales with the length of the chain. This leakage rate $\Gamma_{0}(L)$, in the case of Gaussian distributed white noise, is plotted against chain length in \cref{fig:leakage}. Note that leakage curves for parameters away from the sweet spot (orange and green) should not be trusted for chains of length $L < 5$ (see \cref{app:leakage}).

\Cref{fig:leakage} demonstrates that, when the chains are in the topological phase, the leakage rate is essentially constant with increasing chain length. This suggests that the Rabi oscillations demonstrated in \cref{sec:unitary-dynamics} persist in parameter regimes away from the fine-tuned case simulated in \cref{sec:unitary-dynamics}.

\subsection{\label{sec:two-qubit}Two-qubit gate}

We now show that linking two MT qubits with another electronically controlled topological junction implements a two-qubit gate. To demonstrate this, we create a model for the two-qubit system by extending the single-qubit models of \cref{sec:dynamics}. We then project this system onto the two-qubit subspace. Crucially, the mechanism underlying the two-qubit gate operation is identical to that of the single-qubit gate. This means that the operation speed and the gate fidelity of the two-qubit gate is comparable to the single-qubit case.

\begin{figure}
    \centering
    \includegraphics{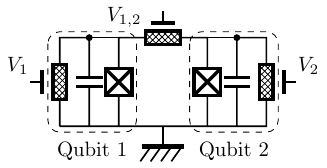}
    \caption{Circuit diagram for the implementation of a two-qubit gate with an inter-qubit junction. The crosshatched lumped elements are the $4\pi$-junctions (these are shown as extended elements in \cref{fig:circuit}a). They are controlled by external gate voltages $V_{j}$. The mechanism of the two-qubit gate is identical to that of the single-qubit gate.}
    \label{fig:2-qubit}
\end{figure}

A circuit diagram for the two-qubit gate setup is shown in \cref{fig:2-qubit}.
The Hamiltonian is
\begin{equation}
\hat{H} = \sum_{j = 1}^{2} \left( \hat{H}_{MT}^{(j)}(\hat{\phi}^{(j)}) + \hat{H}_{KC}^{(j)}(\hat{\phi}^{(j)}) \right) + \hat{H}_{C}^{(12)}.
\end{equation}
Where we now define each of these terms. the Hamiltonian $\hat{H}^{(j)}_{\rm MT}(\hat{\phi}^{(j)})$ has the form of \cref{eq:mt-hamiltonian} and models the single qubit $j$ (we assume both qubits have the same system parameters). The inter-qubit junction is described by $\sum_{j = 1}^{2} \hat{H}_{\rm KC}^{(j)}(\hat{\phi}^{(j)}) + \hat{H}_{\rm C}^{(12)}$ where $\hat{H}_{\rm KC}^{(j)}(\hat{\phi}^{(j)})$ is a Kitaev chain of length $L$ proximity-coupled to the superconducting island of qubit $j$ with phase $\hat{\phi}^{(j)}$,
\begin{align}
\hat{H}_{\rm KC}^{(j)}(\hat{\phi}^{(j)}) = \sum_{k = 1}^{L}\big( -\frac{\mu}{2}{{\hat{c}^{(j)}_{k}}}{}^{\dagger}\hat{c}
_{k}^{(j)} -t {{\hat{c}^{(j)\dagger}_{k}}}{} \hat{c}_{k + 1}^{(j)} \nonumber \\ +  |\Delta|e^{i \hat{\phi}^{(j)}} {\hat{c}_{k + 1}^{(j)\dagger}} {{\hat{c}_{k}^{(j)\dagger}}} + h.c. \big),
\end{align}
where we assume $\mu$, $t$ and $\Delta$ are the same as those of each qubit for simplicity (in contrast to the single-qubit junctions, where only one Kitaev chain Hamiltonian involved the superconducting phase, here, there is a $\hat{\phi}^{(j)}$-dependent term in both inter-qubit chains due to the proximity coupling to the ungrounded superconductor of each qubit). The fermionic annihilation operator $\hat{c}_{k}^{(j)}$ is associated to a fermion on site $k$ of the inter-qubit chain on the side of qubit $j$. Finally, the Hamiltonian describing the inter-qubit coupling junction is
\begin{equation}
\hat{H}_{\rm C}^{(12)} = -w_{1,2}\big( \hat{c}_{L}^{(1)\dagger} \hat{c}_{1}^{(2)} + \hat{c}_{1}^{(2)\dagger} \hat{c}_{L}^{(1)} \big),
\end{equation}
where the real parameter $w_{1,2}$ controls the single-electron tunneling across the inter-qubit junction and is controlled by an external gate voltage $V_{1,2}$.

For the single-qubit system in \cref{sec:topological-transmon}, we performed a change of frame such that the only terms containing couplings between the transmon and the Kitaev chain subsystems appear in the Hamiltonian with the controllable amplitude $w$ as a prefactor. Here, we also perform a change of frame which is effected by a different unitary transformation to that of \cref{sec:topological-transmon}. This unitary transformation is constructed so that the single-qubit Hamiltonians each have the same form as the transformed Hamiltonian $\hmttilde$. The transformation also ensures that the only terms in the inter-qubit junction Hamiltonian which couple the phase operators $\hat{\phi}^{(j)}$ and the Kiteav chains appear with the controllable $w_{1,2}$ amplitude as a prefactor. A unitary with the required properties is given by the operator
\begin{equation}
\hat{U}^{(12)} = \exp \Big[\frac{i}{2} \Big( \big( \hat{n}^{(1)}_{r} + \hat{n}^{(1)}_{KC} \big) \hat{\phi}^{(1)} + \big( \hat{n}^{(2)}_{l} + \hat{n}^{(2)}_{KC} \big)  \hat{\phi}^{(2)} \Big) \Big],
\label{eq:2-qubit-unitary-transformation}
\end{equation}
where the operators $\hat{n}_{l}^{(1)}$ and $\hat{n}_{r}^{(2)}$ are the total charge on the ungrounded single-qubit Kitaev chains and the operator $\hat{n}_{\rm KC}^{(j)}$ is the charge on the inter-qubit chain coupled to qubit $j$, $\hat{n}_{\rm KC}^{(j)} = \sum_{x = 0}^{L - 1} \hat{c}_{x}^{(j)\dagger} \hat{c}_{x}^{(j)}$ (cf. $\rightchaincharge$ in \cref{eq:change-of-frame}). The derivation of how $\hat{H}$ transforms under $\hat{U}^{(12)}$ are given in \cref{app:projection-2-qubits}.

After this transformation, the Hamiltonian is
\begin{equation}
\tilde{H} = \sum_{j = 1}^{2} \left( \tilde{H}_{\rm MT}^{(j)}(\hat{\phi}^{(j)}) + \tilde{H}_{\rm KC}^{(j)} \right) + \tilde{H}_{\rm C}^{(12)}(\hat{\phi}^{(1)},\hat{\phi}^{(2)}).
\label{eq:two-qubit-hamiltonian-after-unitary}
\end{equation}
The transformed single-qubit Hamiltonians $\tilde{H}_{\rm MT}^{(j)}(\hat{\phi}^{(j)})$ take the same form as \cref{eq:hamiltonian-mt-new-basis}, the transformed Hamiltonians $\tilde{H}_{\rm KC}^{(j)}$ are related to $\hat{H}_{\rm KC}^{(j)}$ by the substitution $|\Delta|e^{i \hat{\phi}^{(j)}} \to |\Delta|$, and the transformed inter-qubit junction Hamiltonian is
\begin{align}
\tilde{H}_{\rm C}^{(12)}(\hat{\phi}^{(1)},\hat{\phi}^{(2)}) = & -w_{1,2} \bigl( e^{-i \hat{\phi}^{(12)} / 2} \hat{c}_{L}^{(1)\dagger}\hat{c}_{1}^{(2)} \nonumber \\
&+ e^{i \hat{\phi}^{(12)}/ 2} \hat{c}_{1}^{(2)\dagger} \hat{c}_{L}^{(1)} \bigr),
\end{align}
where $\hat{\phi}^{(12)} \equiv \hat{\phi}^{(2)} - \hat{\phi}^{(1)}$.

Next, we project the Hamiltonian into $\mathcal{H}_{\rm MT}^{(2)}$ to determine how it acts on the qubits defined by the two devices. We perform this projection by making the same simplification we did in \cref{sec:unitary-dynamics}. Namely, we specialise to two fermion sites on each Kitaev chain, $L = 2$, and treat the model at the sweet spot $\mu = 0$, $t = \Delta$. Performing a Jordan-Wigner transformation and projecting onto $\mathcal{H}_{\rm MT}^{(2)}$ (see \cref{app:projection-2-qubits}), we arrive at the $4$-dimensional system
\begin{align}
\tilde{H}_{P} = \sum_{j = 1}^{2} \big( E_{01} \tilde{Z}_{j} + \frac{w_{j}(t)}{2}\cos(\phi^{(j)} / 2)_{10} \tilde{X}_{j} \big) \nonumber \\ 
+ \frac{w_{1,2}(t)}{2} \cos(\phi^{(12)} / 2)_{00,11} \tilde{X}_{1} \tilde{X}_{2},
\label{eq:projected-2-qubit-hamiltonian}
\end{align}
where
\begin{equation}
\cos(\phi^{(12)} / 2)_{00,11} = \bra{\psi_{0},\psi_{0}}\cos(\hat{\phi}^{(12)}/2)\ket{\psi_{1},\psi_{1}}
\end{equation}
and $\tilde{X}_{j},\tilde{Z}_{j}$ are Pauli operators on qubit $j$. We include the time-dependence of $w_{j}(t)$ to demonstrate the protocol for performing two-qubit gates by controlling $w_{1,2}(t)$. 

A protocol for performing an $R_{XX}$ gate using this device is as follows. We initialise the chains in the state
\begin{equation}
\ket{\Omega} = \ket{\Omega}^{(1)} \otimes \ket{\Omega}^{(2)} \otimes \ket{\Omega}^{(12)},
\end{equation}
where $\ket{\Omega}^{(j)}$ has the form of \cref{eq:initial-chain-state} for each of the three junctions, and we take the transmons to be both in their ground states $\ket{\tilde{\psi}_{0}}^{(1)} \otimes \ket{\tilde{\psi}_{0}}^{(2)}$. Next, we send a DC signal through the gate voltage $V_{1,2}$ to control the tunnelling potential $w_{1,2}(t)$ such that
\begin{equation}
w_{1,2}(t) =
\begin{cases}
    0, & t < 0 \\
    w_{1,2}, & 0 < t < t_{\rm gate} \\
    0, & t > t_{\rm gate}.
\end{cases}
\end{equation} 
This DC signal effects a unitary operator on the two-qubit subspace given by
\begin{equation}
\tilde{R}_{XX}(t_{\rm gate}) = \exp \big[ - i \frac{w_{1,2}}{2} \cos(\phi^{(12)} /2)_{00,11} \tilde{X}_{1} \tilde{X}_{2} t_{\rm gate} \big].
\end{equation}
For arbitrary $t_{\rm gate}$, this is an entangling gate on the two MT qubits.

\section{\label{sec:discussion}Discussion}

To summarize, we have shown that a topological superconducting junction realizes the $4\pi$-periodic element identified in Ref \cite{thanh_le_building_2020} as required to perform single- and two-qubit gates on a protected qubit. We demonstrated the gate operation by simulating the dynamics of a minimal Kitaev chain junction. We used this simulation as motivation to analytically project the dynamics of the device onto a qubit subspace and found that single-qubit $R_{X}$-gates and two-qubit $R_{XX}$-gates can be performed by electronically controlling the tunnelling potential across the junction using DC signals. Finally, we used Fermi's golden rule to characterise the effect of charge noise on the operation of the gate. We found that charge noise causes leakage out of the computational subspace, the rate of which is constant in the length of the Kitaev chains. We found that this result persists when parameters are detuned from the sweet spot for the Kitaev chain while remaining in the topological phase. We therefore expect gate operations to be possible for Kitaev chains much longer than the minimal models considered here.

In the context of the fractional Josephson effect, our ground-up simulations demonstrate that the effect does not only manifest in the spectrum of the system but also in its dynamics. This result also provides an alternative perspective on topological-transmon hybrid qubits \cite{hassler_top-transmon_2011, ginossar_microwave_2014}. This perspective treats the topological junction as a circuit element in parallel with the transmon that is $4\pi$-periodic in the superconducting phase difference across the junction. In this perspective, the $4\pi$-periodic topological junction allows access to states of the transmon that are not traditionally accessible \cite{thanh_le_building_2020}.

We highlight that the input to our simulations is the Kitaev chain model with single-electron tunneling across the junction and parameters that are tuned to the topological sweet spot;
the existence of Majorana zero modes at the junction is not 
the starting assumption, although it follows from the
analysis of the model. Nevertheless, our analysis of charge noise 
based on Fermi's golden rule does leverage the existence of
Majorana zero modes.
Further, \cref{fig:leakage} demonstrates that the dynamical effects we observe are robust when the system is detuned from the sweet spot for longer chains, as we expect of an exponentially
localized wavefunction of Majorana zero modes.

As a superconducting circuit element for quantum computation, the MT qubit may find its place in an architecture where qubit control and readout is achieved using only electronically controlled gate voltages. Our results demonstrate that electronically controlled $R_{X}$-gates may be performed on the MT qubit. Further, it has been shown that, electronically controlled $R_{Z}$-gates may be performed in semiconductor-superconductor hybrid systems called gatemons (though RF driving is required, the control mechanism uses DC signals supplied to a gate electrode) \cite{larsen_semiconductor-nanowire-based_2015}. Implementations of topological junctions often include such semiconductor hybrid designs \cite{dvir_realization_2023} and thus, it is conceivable for such devices to implement electronically controlled $R_{X}$- and $R_{Z}$-gates. Such a device would allow for arbitrary single-qubit gates which, along with the $R_{XX}$-gates implemented here, would constitute a universal set of quantum gates on MT qubits. Further, the one- and two-qubits gates would be implimented using similar physical mechanisms and thus have comparable gates speeds (the $R_{Z}$-gates times for gatemons is $\sim 10\,$ns \cite{larsen_semiconductor-nanowire-based_2015}). This is in contrast to traditional two-qubit gates in superconducting circuits (called crossed resonance gates) which have gate speeds that are two orders of magnitude slower than the fastest single-qubit gates \cite{kandala_demonstration_2021}.

Chirolli et al. \cite{chirolli_swap_2022} have recently proposed a device which includes a transmon-type qubit in parallel with a topological junction as a sub-component. In their device, the topological superconductor plays the role of a quantum memory in which a superconducting qubit may be stored through the action of a SWAP gate between the two qubits. This SWAP works as a topological-qubit-to-superconducting-qubit interface. It allows for an architecture where computations are performed on 
easy-to-manipulate superconducting qubits and the outputs are stored in a topological qubit, which acts as a 
resilient memory. Our results suggest that, single- and two-qubit gates may be performed on the superconducting qubit fully electronically without recourse to RF driving.

Further work is needed to determine if such a fully electronic approach to quantum computation with superconducting qubits is viable. In particular, analysis of a device involving both a topological junction and a gatemon would be needed to verify that arbirary single-qubit gates may indeed be implemented fully electronically and with the expected gate speeds. Further, to build a quantum computing architecture which is completely electronically controlled would also require a qubit readout method that does not involves RF sources (RF readout of MT qubits has been demonstrated \cite{yavilberg_fermion_2015}). How this might be achieved is the topic of further work.

Finally, protection of the MT qubit from relaxation is predicated on the idea that local perturbations do not break the $2\pi$-periodicity of the Josephson potential. While this is true when the gate is not operating, we must leave this protected regime in order to couple distinct quasicharge states and perform a gate. Whether the fidelity of the gates is improved by tuning circuit parameters is a subject of future work.

In the context of a transmon in which the spectrum of $\transmonphase$ is real \cite{thanh_le_building_2020}, we are using the topological junction to induce transitions between a state in the centre and a state at the edge of the Brillouin zone. The band structure in \cref{fig:band-structure} tempts one to expect that a mechanism exists to adiabatically vary the Bloch quasicharge $\kappa$ from the middle to the edge of the Brillouin zone and potentially perform a gate without ever leaving the protected regime. The possibility of implementing a protected gate using adiabatic evolution is an interesting open question.

\begin{acknowledgements}
A.A., T.M.S and N.M.C. acknowledge support of the
Australian Research Council Centre
of Excellence for Engineered Quantum Systems 
(Grant No.\ CE170100009). A.A. acknowledges the support of a New Faculty Start-up Grant by Concordia University. N.M.C. would like to acknowledge the support of an Australian Government Research Training Program (RTP) Scholarship doi.org/10.82133/C42F-K220.
\end{acknowledgements}

\bibliography{references.bib}

\appendix

\section{\label{app:unitary-transform}Change of frame for analysis of MT qubit}

We would like to show that, under the change of frame effected by the unitary transformation in \cref{eq:change-of-frame}, the operators composing $\hmt$ transform as
\begin{align}
\hat{\sccharge} \mapsto \hat{\sccharge} - \rightchaincharge/2 
\label{eq:charge-transform}\\
\rightchainfermion{j} \mapsto e^{-i \transmonphase / 2} \rightchainfermion{j}.
\label{eq:fermion-transform}
\end{align}

To prove \cref{eq:charge-transform}, we note that $\rightchaincharge$ commutes with $\hat{\sccharge}$ and $\transmonphase$, and that the canonical commutator between $\hat{\sccharge}$ and $\transmonphase$ implies that, for integer $m \ge 1$,
\begin{equation}
[ \hat{\sccharge}, \transmonphase^{m}] = -im \transmonphase^{m - 1}.
\end{equation}
This implies
\begin{equation}
[\hat{\sccharge}, \unitarytransform] = [\hat{\sccharge}, e^{i \transmonphase \rightchaincharge / 2}] = \frac{\rightchaincharge}{2} e^{i \transmonphase \rightchaincharge / 2} = \frac{\rightchaincharge}{2} \unitarytransform. 
\end{equation}
Therefore,
\begin{equation}
\unitarytransform \hat{\sccharge} \unitarytransform^{\dagger} = \hat{\sccharge} - [\hat{\sccharge}, \unitarytransform]\unitarytransform^{\dagger} = \hat{\sccharge} - \frac{\rightchaincharge}{2},
\end{equation}
as required.

To prove \cref{eq:fermion-transform}, we note that $[\rightchainfermion{j},\transmonphase] = 0$ for all $j$ and that $(\rightchainfermion{j}^{\dagger} \rightchainfermion{j})^{m} = \rightchainfermion{j}^{\dagger} \rightchainfermion{j}$ for integer $m \ge 1$. We will first determine the commutator
\begin{equation}
[\rightchainfermion{j}, \unitarytransform] = [\rightchainfermion{j}, e^{i \transmonphase \rightchaincharge / 2}] = [\rightchainfermion{j}, e^{i \transmonphase \rightchainfermion{j}^{\dagger} \rightchainfermion{j} / 2}] \prod_{i \neq j} e^{i \transmonphase \rightchainfermion{i} \rightchainfermion{i}/2}.
\end{equation}
The commutator on the right hand side is
\begin{equation}
[\rightchainfermion{j}, e^{i \transmonphase \rightchainfermion{j}^{\dagger} \rightchainfermion{j} / 2}]  = [\rightchainfermion{j}, \rightchainfermion{j}^{\dagger} \rightchainfermion{j}] (e^{i \transmonphase /2} - 1) = \rightchainfermion{j} (e^{i \transmonphase /2} - 1).
\end{equation}
Therefore, the transformation of $\rightchainfermion{j}$ is
\begin{align}
\unitarytransform \rightchainfermion{j} \unitarytransform^{\dagger} = \rightchainfermion{j} - [\rightchainfermion{j}, \unitarytransform] \unitarytransform^{\dagger} \nonumber \\
= \rightchainfermion{j} - \rightchainfermion{j} (e^{i \transmonphase /2} - 1) e^{-i \transmonphase \rightchainfermion{j}^{\dagger} \rightchainfermion{j} / 2}.
\end{align}
We then use the fact that $\rightchainfermion{j}e^{-i \transmonphase \rightchainfermion{j}^{\dagger} \rightchainfermion{j} / 2} = \rightchainfermion{j}e^{-i \transmonphase/ 2}$ to obtain the required result.

Finally, note that $\unitarytransform$ commutes with all $\leftchainfermion{j}$ and $\transmonphase$, so the Josephson potential in $\htransmon$ and the left Kitaev chain Hamiltonian $\kitaevhamiltonian{(l)}$ are unaffected by this transformation.

\section{\label{app:jordan-wigner}}

To make the Hamiltonian \cref{eq:hamiltonian-mt-new-basis} easier to simulate, we perform a Jordan-Wigner transform in which we replace our set of fermion sites with an equivalent representation in terms of spins. The general Jordan-Wigner transform of our fermonic operators is \newcommand{\jwspin}[2]{\hat{\sigma}_{#1}^{#2}}
\begin{align}
\rightchainfermion{j} & \to \prod_{k=1}^{j + \chainlength - 1} (\jwspin{k}{z}) \jwspin{j + \chainlength}{-} \\
\leftchainfermion{j} & \to \prod_{k=1}^{j - 1} (\jwspin{k}{z}) \jwspin{j}{-}.
\label{eq:jordan-wigner-transform}
\end{align}
Here, the spin ladder operators are $\jwspin{j}{-} = (\jwspin{j}{x} + i\jwspin{j}{y})/2$ with Pauli matrices $\jwspin{j}{x,y,z}$. The $2L$ spin sites are indexed by $j$ (we keep the hats on spin operators). These forms ensure that the fermion operators obey the canonical anticommutation relations while allowing a natural tensor product structure to be imposed on the Hilbert space (i.e. $\mathcal{H}_{\rm JW} = \bigotimes_{j = 1}^{L} \mathbb{C}_{j}^{2}$).

In the minimal model of the Kitaev chain junction treated in \cref{sec:unitary-dynamics}, we set $L = 2$. It is then simple to explicitly write the transformation of our fermion operators: $\rightchainfermion{1} \mapsto \jwspin{1}{z} \jwspin{2}{z} \jwspin{3}{-}$, $\rightchainfermion{2} \mapsto \jwspin{1}{z} \jwspin{2}{z} \jwspin{3}{z} \jwspin{4}{-}$, $\leftchainfermion{1} \mapsto \jwspin{1}{-}$ and $\leftchainfermion{2} \mapsto \jwspin{1}{z} \jwspin{2}{-}$.

The Hamiltonian of the full system for our choice of parameters, $\chainlength = 2$, $\onsite  = 0$ and $\hopping = |\pairing| \equiv w_{\rm F}$, is given by
\begin{align}
\hmttilde = \htransmontilde(\transmonphase) + w_{\rm F} ( \leftchainfermion{1}^{\dagger} \leftchainfermion{2} - \leftchainfermion{1}^{\dagger} \leftchainfermion{2}^{\dagger} + \rightchainfermion{1}^{\dagger} \rightchainfermion{2} - \rightchainfermion{1}^{\dagger} \rightchainfermion{2}^{\dagger} \nonumber \\
+ \leftchainfermion{2}^{\dagger} \leftchainfermion{1} - \leftchainfermion{2} \leftchainfermion{1} + \rightchainfermion{2}^{\dagger} \rightchainfermion{1} - \rightchainfermion{2} \rightchainfermion{1}) \nonumber \\
- w ( e^{-i \transmonphase / 2} \leftchainfermion{2}^{\dagger} \rightchainfermion{1} + e^{i \transmonphase / 2} \rightchainfermion{1}^{\dagger} \leftchainfermion{2}).
\label{eq:hamiltonian-mt-specific}
\end{align}
Replacing the fermion operators in \cref{eq:hamiltonian-mt-specific} with their spin representations above gives \cref{eq:hamilonian-jordan-wigner}.

\section{\label{app:leakage}Leakage from Fermi's golden rule}

To derive an expression for the leakage out of the computational subspace, we must define the initial and final states $\ket{i} \in \mathcal{H}_{\rm MT}$ and $\ket{f} \in \mathcal{H}_{\rm MT}^{\perp}$ (and so define $\mathcal{H}_{MT}$ itself) to facilitate the computation of the elements $\delta H_{if}$. To do this, note that when $w = 0$ the Hamiltonian $\tilde{H}_{\rm MT}$ given in \cref{eq:hamiltonian-mt-new-basis} is quadratic in bare fermion operators $\rightchainfermion{j}$ and $\leftchainfermion{j}$ and can therefore be cast in terms of quasiparticle operators as\newcommand{\rightchainqp}[1]{\hat{d}_{#1}}\newcommand{\leftchainqp}[1]{\hat{f}_{#1}}
\begin{equation}
    \tilde{H}_{\rm MT}|_{w=0} = \tilde{H}_{\rm T}(\hat{\phi}) + \sum_{j = 0}^{\chainlength - 1} \varepsilon'_{j} \leftchainqp{j}^{\dagger} \leftchainqp{j} + \sum_{j=0}^{\chainlength - 1} \varepsilon_{j} \rightchainqp{j}^{\dagger} \rightchainqp{j},
\end{equation}
where the quasiparticle operators $\leftchainqp{j}$ and $\rightchainqp{j}$ for the left and right chains, respectively, are expressed in terms of bare operators as
\begin{align}
\leftchainqp{j} & = \sum_{x = 1}^{L} ( \beta_{x j}^{*} \leftchainfermion{x} + \varphi_{x j}^{*} \leftchainfermion{x}^{\dagger} ) \nonumber\\
\rightchainqp{j} & = \sum_{x = 1}^{L} ( \alpha_{xj}^{*} \rightchainfermion{x} + \psi_{x j}^{*} \rightchainfermion{x}^{\dagger} )
\label{eq:quasi-particle-in-bare}
\end{align}
(note that, the bare fermion modes are indexed by \mbox{$x = 1, ..., \chainlength$}, while the quasiparticles are indexed by \mbox{$j = 0, ..., L - 1$}). The complex coefficients $\alpha, \beta, \psi$ and $\varphi$ are elements of a pair of $2 \chainlength \times 2 \chainlength$ unitary matrices that are found in practice by diagonalising the Bogoliubov-deGennes Hamiltonian for the system (see \cref{eq:unitary-bogoliubov-left,eq:unitary-bogoliubov-right}). They may always be picked such that the quasiparticle operators $\leftchainqp{j}$ and $\rightchainqp{j}$, obey the canonical anticommutation relations. The coefficients $\alpha_{xj}$, $\psi_{xj}$ may be interpreted as position-space wavefunctions of the fermionic quasiparticles described by $\rightchainqp{j}$ and $\rightchainqp{j}^{\dagger}$, respectively \cite{alase_erasure_2024}. In the topological phase, the energies $\varepsilon_{0}$ and $\varepsilon_{0}'$ vanish \cite{kitaev_unpaired_2001} so the operators $\leftchainqp{0}$ and $\rightchainqp{0}$ describe zero modes on the left and right chains, respectively. For constructing a basis, we define two new zero mode operators as 
\begin{align}
\hat{g}_{0} & = \frac{i}{2} \left( -\rightchainqp{0} + \rightchainqp{0}^{\dagger} + \leftchainqp{0} + \leftchainqp{0}^{\dagger} \right)   \nonumber \\
\hat{h}_{0} & = \frac{i}{2} \left( \rightchainqp{0} + \rightchainqp{0}^{\dagger} - \leftchainqp{0} + \leftchainqp{0}^{\dagger} \right). 
\end{align}
These correspond to the zero modes localised at the junction and at the ends of the Kitaev chains, respectively.

In terms of the bare fermion operators these expand to
\begin{align}
\hat{g}_{0} = \frac{i}{2} \sum_{x = 1}^{L} & \Big( (\psi_{x 0} - \alpha_{x 0}^{*}) \rightchainfermion{x} + (\alpha_{x 0} - \psi_{x 0}^{*}) \rightchainfermion{x}^{\dagger} \nonumber \\
& + (\beta_{x 0}^{*} + \varphi_{x 0})\leftchainfermion{x} + (\beta_{x 0} + \varphi_{x 0}^{*}) \leftchainfermion{x}^{\dagger} \Big) \nonumber \\
\hat{h}_{0} = \frac{i}{2} \sum_{x = 1}^{L} & \Big( ( \alpha_{x 0}^{*} + \psi_{x 0}) \rightchainfermion{x} + (\alpha_{x 0} + \psi_{x 0}^{*}) \rightchainfermion{x}^{\dagger} \nonumber \\
& + (\varphi_{x 0} - \beta_{x 0}^{*} )\leftchainfermion{x} + (\beta_{x 0} - \varphi_{x 0}^{*}) \leftchainfermion{x}^{\dagger} \Big).
\end{align}

We define the Kiteav chain part of the state $\ket{i}$, which we call $\ket{\Omega}$, to be the vacuum for these quasiparticle operators,
\begin{align}
\hat{d}_{j} \ket{\Omega} = \hat{f}_{j} \ket{\Omega} = 0 \quad \text{for } j > 0, \nonumber \\
\hat{g}_{0} \ket{\Omega} = \hat{h}_{0} \ket{\Omega} = 0.
\label{eq:vacuum-definition}
\end{align}
These constraints, along with normalisation, uniquely define $\ket{\Omega}$ on the Kitaev chain subsystems. The state $\ket{\Omega}$ in \cref{eq:initial-chain-state} is a ground state of $\tilde{H}_{\rm JW}$ and an eigenstate of $\tilde{g}_{0}^{\dagger} \tilde{g}_{0}$ with eigenvalue $0$. Therefore, the $\ket{\Omega}$ of \cref{eq:initial-chain-state} is a special case of $\ket{\Omega}$ defined above. Finally, the state of the transmon subsystem in $\ket{i}$ is $\ket{\tilde{\psi}_{i}}$ for $i = 0, 1$. Hence, the span of states $\ket{\tilde{\psi}_{i}} \otimes \ket{\Omega}$ defines $\mathcal{H}_{MT}$.

Next, we take the possible leakage states $\ket{f}$ to have a definite number of quasiparticles,
\begin{equation}
\ket{f;n,m} = 
\begin{cases}
    \hat{d}_{n}^{\dagger} \hat{f}_{m}^{\dagger} \ket{\tilde{\psi}_{f}} \ket{\Omega}, & n \neq 0 \text{ and } m \neq 0 \\
    \hat{d}_{n}^{\dagger} \hat{g}_{0}^{\dagger} \ket{\tilde{\psi}_{f}} \ket{\Omega}, & n \neq 0 \text{ and } m = 0 \\
    \hat{f}_{m}^{\dagger} \hat{g}_{0}^{\dagger} \ket{\tilde{\psi}_{f}} \ket{\Omega}, & n = 0 \text{ and } m \neq 0 \\
    \ket{\tilde{\psi}_{f'}}\ket{\Omega}, & n = 0 \text{ and } m = 0. 
\end{cases}
\label{eq:final-state}
\end{equation}
Here, each final state is specified by a tuple $(f,n,m)$ where $n$ and $m$ specify the two quasiparticle modes excited on each Kitaev chain and range from $0$ to $L - 1$. The possible final states are restricted in anticipation of the fact that each term in $\delta \tilde{H}$ contains two bare fermion operators: one on each chain. This implies that, at first order in perturbation theory, $\delta \tilde{H}$ only induces transitions to states with exactly two quasiparticle excitations and these two excitations cannot be on the same chain (so $\hat{d}_{n}^{\dagger} \hat{g}_{0}^{\dagger} \ket{\Omega}$ is a possible leakage state while $\hat{d}_{n}^{\dagger} \hat{d}_{\ell}^{\dagger} \ket{\Omega}$ is not). The only leakage states which have no quasiparticle excitations of the Kitaev chains, must have the transmon subsystem in a state $\ket{\tilde{\psi}_{f'}}$ for $f' > 1$. Further, we have omitted states with an excitation of the quasiparticle described by $\hat{h}_{0}^{\dagger} \hat{h}_{0}$ as will be justified below. The transmon state $\ket{\tilde{\psi}_{f}}$ can be any eigenstate of $\tilde{H}_{\rm T}(\hat{\phi})$. Leakage rates into states of a form other than \cref{eq:final-state} vanish.

Finally, we invert the expansion \cref{eq:quasi-particle-in-bare} to express the bare operators appearing in $\delta \tilde{H}(t)$ in terms of quasiparticles
\begin{align}
\rightchainfermion{1} = \sum_{x = 1}^{\chainlength - 1} ( \alpha_{1, x} \rightchainqp{x} & + \psi_{1, x}^{*} \rightchainqp{x}^{\dagger} ) \nonumber \\ 
 & + \frac{i}{2} ( \psi_{1, 0}^{*} - \alpha_{1, 0}) ( \hat{g}_{0} + \hat{g}_{0}^{\dagger} )\nonumber \\
\leftchainfermion{L} = \sum_{x = 1}^{\chainlength - 1} ( \beta_{\chainlength, x} \leftchainqp{x} & + \varphi_{\chainlength, x} ^{*} \hat{f}_{x}^{\dagger} ) \nonumber \\
 & + \frac{i}{2} ( \varphi_{\chainlength , 0}^{*}  + \beta_{\chainlength, 0}) ( \hat{g}_{0}^{\dagger} - \hat{g}_{0}) .
\label{eq:bare-in-quasi-particles}
\end{align}
We have assumed that the chain length is much larger than the Majorana localisation length which implies that the wavefunctions of operators $\hat{h}_{0}$ corresponding to the zero modes local to the ends of the wire are negligible at the junction. This allows us to drop the factor of $\hat{h}_{0}$ in \cref{eq:final-state} as it commutes with the perturbation $\delta \tilde{H}$. This assumption is why the leakage curves for parameters away from the sweet spot in \cref{fig:leakage} should not be trusted for short chains.

We substitute the expressions for $\hat{a}_{1}$ and $\hat{b}_{L}$ into the \eqref{eq:hamiltonian-perturbing} and compute the elements $\delta H^{nm}_{if}$. We arrive at an expression for these elements,
\begin{equation}
\delta H_{if}^{nm} = \begin{cases}
A_{0 0} (e^{i \phi / 2})_{i f'}  + B_{00} (e^{-i \phi / 2})_{i f'}, & n, m = 0\\
A_{n m} (e^{i \phi / 2})_{i f} + B_{n m} (e^{-i \phi / 2})_{i f}, &  \text{otherwise.}

\end{cases}
\end{equation}
where $(e^{\pm i \phi / 2})_{i f} = \bra{\psi_{i}} e^{\pm i \hat{\phi} / 2}\ket{\psi_{f}}$,
\begin{equation}
A_{n m} = 
\begin{cases}
\alpha_{1, n}^{*} \varphi_{\chainlength, m}^{*}, & n \neq 0, m \neq 0 \nonumber \\
\frac{i}{2} ( \varphi_{\chainlength, 0}^{*} + \beta_{\chainlength, 0} ) \alpha_{1, n}^{*}, & n \neq 0, m = 0 \\
\frac{i}{2} (\psi_{1, 0} - \alpha_{1, 0}^{*}) \varphi_{\chainlength, m}^{*} & n = 0, m \neq 0 \\
-\frac{1}{4} (\psi_{1, 0} - \alpha_{1, 0}^{*}) ( \varphi_{\chainlength, 0}^{*} + \beta_{\chainlength, 0}), & n,m = 0
\end{cases}
\end{equation}
and,
\begin{equation}
B_{n m} =
\begin{cases}
-\beta_{\chainlength, m}^{*} \psi_{1, n}^{*}, & n \neq 0, m \neq 0 \\
-\frac{i}{2} ( \varphi_{\chainlength, 0} + \beta_{\chainlength, 0}^{*}) \psi_{1, n}^{*}, & n \neq 0, m = 0 \\
\frac{i}{2} ( \psi_{1, 0}^{*} - \alpha_{1, 0}) \beta_{\chainlength, m}^{*} , & n = 0, m \neq 0 \\
-\frac{1}{4} (\psi_{1, 0}^{*} - \alpha_{1, 0}) ( \varphi_{\chainlength, 0} + \beta_{\chainlength, 0}^{*}), & n,m = 0.
\end{cases}
\end{equation}
The leakage rate out of the qubit subspace $\mathcal{H}_{MT}$ into the state $\ket{f; n,m}$ is then
\begin{equation}
\Gamma_{if}^{nm} = 2\pi \left| \delta H_{if}^{nm} \right|^{2} S(E_{f} - E_{i}).
\end{equation}
The total leakage rate out of the state $\ket{i}$ is then found by summing over all final states $\ket{f;n,m}$,
\begin{multline}
\Gamma_{i}(L) = \sum_{f,n,m}\Gamma_{if}^{nm} =  \\
2 \pi \sum_{f = 0}^{\infty} \sum_{n,m = 0}^{L - 1} S(E_{f}) |(e^{i \phi /2})_{if}A_{nm} + (e^{-i \phi /2})_{if}B_{nm}|^{2} \\
- 2\pi \sum_{f' = 0}^{1} S(E_{f'}) | (e^{i \phi /2})_{i f'} A_{0 0} + (e^{-i \phi /2})_{i f'} B_{0 0}  |^{2}
\label{eq:leakage-full}
\end{multline}

Finally, to produce the plot of $\Gamma_{0}(L)$ in \cref{fig:leakage}, we numerically diagonalise the Bogoliubov-deGennes Hamiltonian corresponding to $\tilde{H}_{\rm MT}|_{w=0}$, expressed in block form as a $2 \chainlength \times 2 \chainlength$ Hermitian matrix,
\begin{equation}
\mathbf{H}_{BdG} =
    \begin{pmatrix}
    \mathbf{A} & \mathbf{B}^{*} \\
    -\mathbf{B} & -\mathbf{A}^{*}
    \end{pmatrix}
= \mathbf{U}_{BdG} \mathbf{D} \mathbf{U}_{BdG}^{\dagger}.
\end{equation}
where $\mathbf{A} = \mathbf{A}^{\dagger}$ and $\mathbf{B} = - \mathbf{B}^{T}$ are $L \times L$ matrices, $\mathbf{U}_{BdG}$ is unitary and $\mathbf{D}$ is diagonal \cite{van_hemmen_note_1980}. The coefficients $\alpha_{jx}$ and $\psi_{jx}$ are then related to the entries of the matrix $\mathbf{U}^{r}_{BdG}$ that diagonalises the Bogoliubov-deGennes Hamiltonian corresponding to the right chain as
\begin{equation}
\mathbf{U}^{r}_{BdG} = 
\begin{pmatrix}
    \bm{\alpha} & \bm{\psi}^{*} \\
    \bm{\psi} & \bm{\alpha}^{*}
\end{pmatrix}.
\label{eq:unitary-bogoliubov-left}
\end{equation}
Where $\bm{\alpha}$ has elements $\alpha_{ij}$ and $\bm{\psi}$ has elements $\psi_{ij}$. The other coefficients are given in terms of the corresponding unitary for the left chain as
\begin{equation}
\mathbf{U}^{l}_{BdG} = 
\begin{pmatrix}
    \bm{\beta} & \bm{\varphi}^{*} \\
    \bm{\varphi} & \bm{\beta}^{*}
\end{pmatrix}.
\label{eq:unitary-bogoliubov-right}
\end{equation}
Where $\bm{\varphi}$ has elements $\varphi_{i j}$ and $\bm{\beta}$ has elements $\beta_{i j}$.

\section{Projection onto the qubit subspace}

\subsection{\label{app:projection}Projection onto the single-qubit subspace}

We wish to project the Hamiltonian of the full, single-qubit system $\tilde{H}_{\rm MT}$, given in \cref{eq:hamilonian-jordan-wigner}, onto the qubit subspace $\mathcal{H}_{\rm MT}$, spanned by states $\ket{\tilde{\psi}_{0}} \otimes \ket{\Omega}$ and $\ket{\tilde{\psi_{1}}} \otimes \ket{\Omega}$. To perform this projection, we notice several facts. Fact 1: The state $\ket{\Omega}$ is an eigenvector of the operators $\hat{\sigma}_{1}^{x} \hat{\sigma}^{x}_{2}$ and $\hat{\sigma}_{3}^{x} \hat{\sigma}_{4}^{x}$ with eigenvalue $1$. Fact 2: The matrix element $\bra{\Omega} \hat{\sigma}_{2}^{+} \hat{\sigma}_{3}^{-} \ket{\Omega} = 1/4$ which implies $\bra{\Omega} \hat{\sigma}_{3}^{+} \hat{\sigma}_{2}^{-} \ket{\Omega} = 1/4$. Fact 3: The expectation value of $\cos(\hat{\phi}/2)$ in any eigenstate $\ket{\tilde{\psi}_{i}}$ of $\tilde{H}_{\rm T}$ vanishes. The first two facts are easily seen by a direct computation. Fact 3 is proved by expanding the inner product as an integral over the compact phase interval $-2\pi$ to $2\pi$ (recall that we have expanded the Hilbert space to accommodate $4\pi$-periodic functions of $\phi$). The fact that $\cos(\phi/2)$ is $2\pi$ antiperiodic while $|\psi_{i}(\phi)|^{2}$ is $2\pi$ periodic shows that the inner product vanishes.

Using facts 1 and 2 above, it is easy to show that the matrix elements of the $2 \times 2$ matrix $\tilde{H}_{\rm MT}^{(P)}$ are
\begin{equation}
\bra{\psi_{i}}\bra{\Omega} \tilde{H}_{\rm MT} \ket{\psi_{j}} \ket{\Omega}= (E_{i} + w_{F})\delta_{ij} + w\cos(\phi/2)_{ij}/2.
\label{eq:single-qubit-hamiltonian-qubit-subspace}
\end{equation}
By Fact 3, $\cos(\phi/2)_{ij}$ vanishes for $i = j$. Therefore, shifting \cref{eq:single-qubit-hamiltonian-qubit-subspace} by the constant $(w_{F} - (E_{0} + E_{1})/2 )\delta_{ij}$ implies equation \cref{eq:hamiltonian-effective-theory}.

\subsection{\label{app:projection-2-qubits}Projection onto the two-qubit subspace}

We wish to accomplish two things in this appendix. Firstly, we transform $\hat{H}$ in the original picture into $\tilde{H}$ in the new picture by applying the unitary transformation $\hat{U}^{(12)}$ given in \cref{eq:2-qubit-unitary-transformation}. Secondly, we project $\tilde{H}$ of \cref{eq:two-qubit-hamiltonian-after-unitary} into the two-qubit subspace $\mathcal{H}_{\rm MT}^{(2)}$, which we will properly define below. 

The transformation $\hat{U}^{(12)}$ affects each operator in the Hamiltonian $\hat{H}$ as follows. The transmon charge operators for qubit 1 is shifted as 
\begin{equation}
\hat{n}^{(1)} \mapsto \hat{n}^{(1)} - (\hat{n}_{r}^{(1)} + \hat{n}_{\rm KG}^{(1)})/2,
\end{equation}
and the transmon charge operator for qubit 2 is shifted as
\begin{equation}
\hat{n}^{(2)} \mapsto \hat{n}^{(2)} - (\hat{n}_{l}^{(2)} + \hat{n}_{\rm KG}^{(2)}) / 2.
\end{equation}
The fermion operators $\hat{a}_{k}^{(j)}$ for each ungrounded single-qubit chain each pickup a $e^{-i \hat{\phi}^{(j)} / 2}$ factor as in the single-qubit case: $\hat{a}_{k}^{(j)} \mapsto e^{-i \hat{\phi}^{(j)} / 2} \hat{a}_{k}^{(j)}$. Similarly, the fermion operators on the inter-qubit chains $\hat{c}_{k}^{(j)}$ transform as $\hat{c}_{k}^{(j)} \mapsto e^{-i \hat{\phi}^{(j)}}\hat{c}_{k}^{(j)}$ (here, the fermion operators on both chains pick up the $\hat{\phi}^{(j)}$-dependent factor as neither is grounded). The derivation of each of these transformations follows the same steps as the ones outlined for the single-qubit transformation in \cref{app:unitary-transform}.

We define the subspace $\mathcal{H}_{\rm MT}^{(2)}$ as
\begin{equation}
    \mathcal{H}_{\rm MT}^{(2)} \equiv \text{span}\{\ket{\tilde{\psi}_{i}}^{(1)}\ket{\tilde{\psi}_{j}}^{(2)}\ket{\Omega}, i,j = 0, 1\}.
\end{equation}
Where $\ket{\tilde{\psi}_{i}}^{(j)}$ are eigenstates of the transmon part of the Hamiltonian for each single-qubit $\tilde{H}_{\rm T}^{(j)}$ and \mbox{$\ket{\Omega}\equiv \ket{\Omega}^{(1)}\ket{\Omega}^{(2)}\ket{\Omega}^{(12)}$} where $\ket{\Omega}^{(j)}$ is a ground state of each of the junctions of the same form as \cref{eq:initial-chain-state}.

We specialise to $\chainlength = 2$, $\onsite = 0$ and $\hopping = |\pairing|$, as in the single-qubit case. We then perform a Jordan-Wigner transformation in the same way as in \cref{sec:unitary-dynamics}. In this case, there are $12$ spin sites indexed by $j$: four on each of the single-qubit Kitaev chain junctions and four on the inter-qubit junction. We will take spin operators with index $j$ between $5$ and $8$ to correspond to the inter-qubit chain. In particular, the operators appearing in the coupling junction Hamiltonian $\tilde{H}_{\rm C}^{(12)} (\hat{\phi}^{(1)}, \hat{\phi}^{(2)})$ are $\jwspin{6}{\pm}$ and $\jwspin{7}{\pm}$.

We are now ready to project $\tilde{H}$ onto the qubit subspace. First, projecting the single-qubit Hamiltonians $\sum_{j = 1}^{2} H_{\rm MT}^{(j)}$ onto this subspace gives the block matrix
\begin{equation}
    \sum_{j = 1}^{2} \tilde{H}_{MT}^{(j)} \to 
    \begin{pmatrix}
        \tilde{H}_{\rm MT}^{P(1)} & \mathbf{0}_{2} \\
        \mathbf{0}_{2} & \tilde{H}_{\rm MT}^{P(2)}.
    \end{pmatrix}
\end{equation}
Where $\tilde{H}_{\rm MT}^{P(j)}$ is a $2 \times 2$ matrix of the same form as \cref{eq:projected-mt-hamiltonian}.

Next, just as in the single-qubit case, the Hamiltonian $\sum_{j = 1}^{2} \tilde{H}_{\rm KC}^{(j)}$, corresponding to the coupling inter-qubit Kitaev chains (excluding the junction term), is an operator proportional to the identity in $\mathcal{H}_{\rm MT}^{(2)}$. We remove this operator by shifting the Hamiltonian by a constant.

Finally, we project the inter-qubit junction Hamiltonian $\tilde{H}_{\rm C}^{(12)}(\hat{\phi}^{(1)}, \hat{\phi}^{(2)})$ into $\mathcal{H}_{\rm MT}^{(2)}$. As in the single-qubit case, matrix elements of the spin operators are \mbox{$\bra{\Omega}\hat{\sigma}_{6}^{+} \sigma_{5}^{-} \ket{\Omega} = 1/4 \nonumber$} and $\bra{\Omega}\hat{\sigma}_{5}^{+} \sigma_{6}^{-} \ket{\Omega} = 1/4$. The projection of $\tilde{H}_{C}^{(12)}$ is therefore
\begin{align}
\bra{\psi_{i}} \bra{\psi_{n}} \tilde{H}_{\rm C}^{(12)} \ket{\psi_{j}} \ket{\psi_{m}} = \nonumber \\ w_{C}\bra{\psi_{i}} \bra{\psi_{n}} \cos((\hat{\phi}^{(2)} - \hat{\phi}^{(1)}) / 2) \ket{\psi_{j}} \ket{\psi_{m}} / 2.
\end{align}
Where $i, j, n$ and $m$ are $0$ or $1$. An extension of the arguments in \cref{app:projection}, shows that the matrix elements \mbox{$\bra{\psi_{i}} \bra{\psi_{n}} \cos((\hat{\phi}^{(2)} - \hat{\phi}^{(1)}) / 2) \ket{\psi_{j}} \ket{\psi_{m}}$} vanish when $i = j$ or $n = m$. Therefore, the only non-vanishing elements of this $4 \times 4$ matrix are on the anti-diagonal. In fact, all of the elements on the anti-diagonal must be equal. To see this, we write those elements as
\begin{align}
 \bra{\psi_{i}} \bra{\psi_{n}} \cos((\hat{\phi}^{(2)} - \hat{\phi}^{(1)}) / 2) \ket{\psi_{j}} \ket{\psi_{m}} \nonumber \\
 = w_{C}\braket{\psi_{j} | e^{i\phi^{(1)}/2} | \psi_{n}}\braket{\psi_{k} | e^{-i\phi^{(2)}/2} | \psi_{m}}/4 \nonumber \\
 + w_{C}\braket{\psi_{j} | e^{-i\phi^{(1)}/2} | \psi_{n}}\braket{\psi_{k} | e^{i\phi^{(2)}/2} | \psi_{m}}/4.
\end{align}
Then, using the facts that $i \neq j$, $n \neq m$ and that the eigenfunctions $\psi_{0}^{(j)}(\phi)$ and $\psi_{1}^{(j)}(\phi)$ are all real, we see that every non-vanishing element is equal. We will pick as a representative element,
\begin{align}
 \bra{\psi_{1}} \bra{\psi_{1}} \cos((\hat{\phi}^{(2)} - \hat{\phi}^{(1)}) / 2) \ket{\psi_{0}} \ket{\psi_{0}} \nonumber \\
 = w_{C}\braket{\psi_{1} | e^{i\phi^{(1)}/2} | \psi_{0}}\braket{\psi_{1} | e^{-i\phi^{(2)}/2} | \psi_{0}}/4 \nonumber \\
 + w_{C}\braket{\psi_{1} | e^{-i\phi^{(1)}/2} | \psi_{0}}\braket{\psi_{1} | e^{i\phi^{(2)}/2} | \psi_{0}}/4.
\end{align}
Hence, the projection of $\tilde{H}_{\rm C}^{(12)}(\hat{\phi}^{(1)}, \hat{\phi}^{(2)})$ onto $\mathcal{H}_{\rm MT}^{(2)}$ is therefore
\begin{align}
{H_{\rm C}^{(P)}}(\hat{\phi}^{(1)}, \hat{\phi}^{(2)}) = &\nonumber\\
(w_{1,2}/2) \bra{\psi_{1}} \bra{\psi_{1}} & \cos((\hat{\phi}^{(2)} - \hat{\phi}^{(1)}) / 2) \ket{\psi_{0}} \ket{\psi_{0}} \tilde{X}_{1}\tilde{X}_{2}.
\end{align}
Where $\tilde{X}_{j}$ is a Pauli-$X$ on qubit $j$. 

Finally, putting all of these results together gives the projected Hamiltonian $\tilde{H}_{P}$ in \cref{eq:projected-2-qubit-hamiltonian}.

\end{document}